\documentclass[twocolumn]{aastex631}

\usepackage[utf8]{inputenc}
\usepackage[T1]{fontenc}
\usepackage{amsmath}
\usepackage{graphicx}

\renewcommand{\vec}[1]{\boldsymbol{#1}}

\defcitealias{thomas_cetus-palca_2022}{TB22}

\graphicspath{{./}{figures/}}

\received{August 11, 2024}
\revised{November 19, 2024}
\accepted{December 6, 2024}

\shorttitle{The Cetus-Palca Stream \& Segue 2}
\shortauthors{Foote et al.}

\begin{document}

\title{Segue 2 Recently Collided with the Cetus-Palca Stream: New Opportunities to Constrain Dark Matter in an Ultra-Faint Dwarf}

\correspondingauthor{Hayden R. Foote}
\email{haydenfoote@arizona.edu}

\author[0000-0003-1183-701X]{Hayden R. Foote}
\affil{Steward Observatory, University of Arizona, 933 North Cherry Avenue, Tucson, AZ 85721, USA}

\author[0000-0003-0715-2173]{Gurtina Besla}
\affil{Steward Observatory, University of Arizona, 933 North Cherry Avenue, Tucson, AZ 85721, USA}

\author[0000-0001-7107-1744]{Nicol\'as Garavito-Camargo}
\affil{Center for Computational Astrophysics, Flatiron Institute, Simons Foundation, 162 Fifth Avenue, New York, NY 10010, USA}

\author[0000-0002-9820-1219]{Ekta Patel}\thanks{NASA Hubble Fellow}
\affil{Department of Physics and Astronomy, University of Utah, 115 South 1400 East, Salt Lake City, UT 84112, USA}

\author[0000-0002-2468-5521]{Guillaume F. Thomas}
\affil{Departamento de Astrofísica, Universidad de La Laguna, E-38206 La Laguna, Tenerife, Spain}
\affil{Instituto de Astrofísica de Canarias, E-38200 La Laguna, Tenerife, Spain}

\author[0000-0002-7846-9787]{Ana Bonaca}
\affil{The Observatories of the Carnegie Institution for Science, 813 Santa Barbara Street, Pasadena, 91101, CA, USA}

\author[0000-0003-0872-7098]{Adrian M. Price-Whelan}
\affil{Center for Computational Astrophysics, Flatiron Institute, Simons Foundation, 162 Fifth Avenue, New York, NY 10010, USA}

\author[0000-0002-8040-6785]{Annika H. G. Peter}
\affil{Department of Astronomy, The Ohio State University, 140 West 18th Avenue, Columbus, OH 43210, USA}
\affil{Center for Cosmology and Astroparticle Physics, The Ohio State University, 191 West Woodruff Avenue, Columbus, OH 43210, USA}
\affil{Department of Physics, The Ohio State University, 191 West Woodruff Avenue, Columbus, OH, 43210, USA}
\affil{School of Natural Sciences, Institute for Advanced Study, 1 Einstein Drive, Princeton, NJ 08540, USA}

\author[0000-0002-5177-727X]{Dennis Zaritsky}
\affil{Steward Observatory, University of Arizona, 933 North Cherry Avenue, Tucson, AZ 85721, USA}

\author[0000-0002-1590-8551]{Charlie Conroy}
\affil{Center for Astrophysics | Harvard \& Smithsonian, 60 Garden Street, Cambridge, MA 02138, USA}



\begin{abstract}
Stellar streams in the Milky Way are promising detectors of low-mass dark matter (DM) subhalos predicted by $\Lambda$CDM. Passing subhalos induce perturbations in streams that indicate the presence of the subhalos.
Understanding how known DM-dominated satellites impact streams is a crucial step towards using stream perturbations to constrain the properties of dark perturbers. Here, we cross-match a \textit{Gaia} EDR3 and SEGUE member catalog of the Cetus-Palca stream (CPS) with H3 for additional radial velocity measurements and fit the orbit of the CPS using this 6-D data. 
We demonstrate for the first time that the ultra-faint dwarf Segue 2 had a recent (77$\pm$5 Myr ago) close flyby (within the stream's 2$\sigma$ width) with the CPS. 
This interaction enables
constraints on Segue 2's mass and density profile at larger radii ($\mathcal{O}(1)$ kpc) than are probed by its stars ($\mathcal{O}(10)$ pc). 
While Segue 2 is not expected to strongly affect the portion of the stream covered by our 6-D data, 
we predict that if Segue 2's mass within $\sim 6$ kpc is $5\times 10^9\,M_\odot$, the CPS's velocity dispersion will be $\sim 40$ km s$^{-1}$ larger at $\phi_1>20^\circ$ than at $\phi_1<0^\circ$.
If no such heating is detected, Segue 2's mass cannot exceed $10^9\,M_\odot$ within $\sim 6$ kpc.
The proper motion distribution of the CPS near the impact site is mildly sensitive to the shape of Segue 2's density profile.
This study presents a critical test for 
frameworks designed to constrain properties of dark subhalos from stream perturbations.

\end{abstract}

\keywords{Stellar streams (2166), Dark matter (353), Dwarf galaxies (416), Milky Way dynamics (1051)}


\section{Introduction} \label{sec:intro}

Stellar streams are formed from the tidal disruption of dwarf galaxies and globular clusters as they orbit within the gravitational potentials of their host galaxies. In the Milky Way (MW), streams offer one of the most promising avenues for constraining the nature of dark matter (DM) at small scales, as they contain signatures of past encounters with the low-mass subhalos predicted by $\Lambda$CDM theory (e.g., \citealt{ibata_uncovering_2002, johnston_how_2002, siegal-gaskins_signatures_2008}; see also \citealt{bonaca_stellar_2025} for a recent review). 

The close passage of a subhalo imparts a gravitational kick on stream stars in the vicinity of the flyby, which leads to the formation of observable perturbations including gaps, kinks, spurs, asymmetries, and/or larger velocity dispersions \citep[e.g.,][]{carlberg_star_2009, carlberg_dynamics_2013,  carlberg_star_2024, yoon_clumpy_2011, erkal_forensics_2015, erkal_properties_2015, ngan_simulating_2015, ngan_dispersal_2016, helmi_time_2016, sanders_dynamics_2016, sandford_quantifying_2017, koppelman_time_2021}. Gaps that are consistent with perturbations from compact substructures have been identified in the GD-1 \citep{bonaca_spur_2019, bonaca_high-resolution_2020} and Pal-5 streams \citep{carlberg_pal_2012, erkal_sharper_2017}. In the case of GD-1, \citet{doke_probability_2022} found that interactions with known globular clusters were unable to reproduce the gaps in the stream, providing evidence for a dark subhalo formation channel.

An important step towards using streams as detectors of dark substructure is to study perturbations on streams from \textit{known} substructures, such as MW satellites with constrained 3D velocities and positions. In such cases, the timing and geometry of the encounter, as well as some properties of the perturber, are constrained. Such cases present critical tests of the frameworks being developed to infer the existence and properties of truly dark subhalos. 

In this paper, we report that a close encounter between the Cetus-Palca stream (CPS) and the ultra-faint dwarf (UFD) Segue 2  
must have occurred within the past 100 Myr. We further study the imprints of this encounter on the structure of stellar members of the CPS and how these imprints can constrain the DM density profile and total mass of Segue 2. 

Constraints on the distribution of DM within the Segue 2 UFD would shed light on several open questions in DM physics and the faint end of galaxy formation. 
For example, a key prediction of DM-only cosmological simulations in the $\Lambda$CDM paradigm is that all CDM halos, regardless of mass, will relax into a density profile with a central cusp \citep[e.g.,][]{navarro_diversity_2010}. However, observations of low-mass galaxy rotation curves frequently imply cored density profiles, creating the ``core-cusp problem'' (e.g., \citealt{flores_observational_1994, moore_evidence_1994}; reviews by \citealt{de_blok_core-cusp_2010, bullock_small-scale_2017, salucci_distribution_2019, sales_baryonic_2022} and references therein).\footnote{Though Draco is a notable exception with a DM cusp \citep[e.g.,][]{read_case_2018, hayashi_diversity_2020, vitral_hstpromo_2024}.} Proposed solutions to the core-cusp problem include unaccounted for uncertainties in modeling the internal kinematics of dwarf galaxies \citep[e.g.,][]{strigari_dynamical_2017, genina_core-cusp_2018, harvey_impact_2018, oman_non-circular_2019, santos-santos_baryonic_2020, downing_many_2023, roper_diversity_2023}, supernova feedback \citep[e.g.,][]{dekel_origin_1986, navarro_cores_1996, gnedin_maximum_2002, read_mass_2005, mashchenko_stellar_2008, pontzen_how_2012, 
teyssier_cusp-core_2013, di_cintio_dependence_2014, chan_impact_2015, dutton_nihao_ix_2016, tollet_nihao_2016, fitts_fire_2017, lazar_dark_2020, jahn_real_2023, jackson_formation_2024}, and alternative models for DM, such as self-interacting DM \citep[e.g.,][]{spergel_observational_2000, yoshida_weakly_2000, vogelsberger_subhaloes_2012, rocha_cosmological_2013, tulin_dark_2018, burger_nature_2019, burger_supernova-driven_2021, burger_degeneracies_2022} and fuzzy DM \citep[e.g.,][]{hu_fuzzy_2000, schive_cosmic_2014, marsh_axion_2015, calabrese_ultra-light_2016, chen_jeans_2017, gonzalez-morales_unbiased_2017, safarzadeh_ultra-light_2020}. 

The Segue 2 - CPS interaction is especially useful for disentangling these proposed solutions to the core-cusp problem. Constraints on Segue 2's DM distribution from its effect on the CPS would not be subject to the same modeling uncertainties as Segue 2's internal dynamics. Meanwhile, supernova feedback is predicted to be inefficient at forming cores in UFDs \citep[e.g.,][]{tollet_nihao_2016, fitts_fire_2017}, so constraining the DM density profile of a UFD, such as Segue 2, would provide a critical test of core formation in a regime where the effects of DM particle physics can be separated from the effects of baryonic processes. In addition, a constraint on the total DM mass of Segue 2 would help to understand the halo mass floor needed to host a galaxy \citep[e.g.,][]{wheeler_sweating_2015, jeon_connecting_2017} in $\Lambda$CDM. 

This paper is structured as follows. In Section \ref{sec:target} we provide information on the two targets of this study, the CPS and Segue 2. In Section \ref{sec:obs}, we identify a higher-purity subset of the CPS catalog from \citet{thomas_cetus-palca_2022} (hereafter \citetalias{thomas_cetus-palca_2022}), which we cross-match with the H3 survey \citep{conroy_mapping_2019} to obtain additional radial velocity measurements. In Section \ref{sec:orbit}, we use these stars to fit the orbit of the CPS. We then characterize Segue 2's interaction with the CPS in Section \ref{sec:segue2}, making predictions for the observability of the perturbation Segue 2 leaves in the CPS as a function of the mass and density profile of its DM halo. Section \ref{sec:disc} contains a discussion of the limitations of our models, as well as further exploration of a serendipitous kinematic association of stars in our data. Lastly, we summarize our results and conclude in Section \ref{sec:con}.

\section{Targets: the Cetus-Palca Stream and the Segue 2 UFD}
\label{sec:target}

There is a small but growing body of literature which explores the influence of known MW satellites and globular clusters on stellar streams. For example, the MW's largest satellite, the LMC, can induce track / velocity misalignment in stellar streams \citep[e.g.,][]{erkal_modelling_2018, erkal_total_2019, koposov_piercing_2019, koposov_s5_2023, ji_kinematics_2021, shipp_measuring_2021, vasiliev_tango_2021}, and cause streams to deform in the time-dependent MW+LMC potential \citep{lilleengen_effect_2023, brooks_action_2024}. Additionally, the Sagittarius (Sgr) dwarf galaxy has been proposed as the culprit behind the kink observed in the ATLAS-Aliqa Uma stream \citep{li_broken_2021} as well as the multicomponent appearance of the Jhelum stream \citep{woudenberg_characterization_2023}. Other studies have predicted that the MW's satellite galaxies are expected to affect the orbits and tidal tails of its globular clusters \citep{garrow_effects_2020, el-falou_effect_2022}.

Inspired by these works, we searched for potential interactions between MW satellites and known streams by comparing the \texttt{galstreams} library of MW stellar streams \citep{mateu_galstreams_2023} to a state-of-the-art library of MW satellite orbits by Patel \& Garavito-Camargo et al. (2025, in preparation), finding that the CPS and Segue 2 likely experienced a recent interaction. In this section, we provide further background on these two targets, before confirming and studying their interaction in the remaining sections.

\subsection{The Cetus-Palca Stream}

The Cetus stream was originally identified in the SEGUE survey \citep{yanny_segue_2009} as a group of blue horizontal branch (BHB) stars near the trailing arm of the Sgr stream with metallicities and radial velocities distinct from Sgr \citep{newberg_discovery_2009, yanny_tracing_2009}. \citet{newberg_discovery_2009} also found that the globular cluster NGC 5824 is kinematically associated with the Cetus stream. \citet{yam_update_2013} first modeled Cetus with \textit{N}-body simulations, showing it could be formed by the disruption of a $10^8\,M_\odot$ dwarf galaxy. Motivated by this finding and the stream's association with NGC 5824, \citet{yuan_revealing_2019} investigated whether the cluster might be the stripped core of the Cetus stream's progenitor, strengthening its kinematic association with Cetus and discovering a northern extension to the stream. Later, detailed \textit{N}-body models by \citet{chang_is_2020} demonstrated that NGC 5824 was unlikely to be the Cetus stream's progenitor, as its orbit could not reproduce the ``main'' southern component of the Cetus stream. 

\citet{chang_is_2020} also first associated Cetus with the Palca stream, which had been discovered in Dark Energy Survey data by \citet{shipp_stellar_2018} and further studied by \citet{li_broken_2021} and \citet{li_s_2022}. Subsequently, \citet{yuan_complexity_2022} and \citetalias{thomas_cetus-palca_2022} extended previous detections of the Cetus stream to meet Palca, independently confirming that the Cetus and Palca streams are indeed one structure. Both works suggested that the main part of the Cetus stream be renamed to Cetus-Palca. \citet{yuan_complexity_2022} also identified two additional structures consistent with other wraps of the CPS in the \citet{chang_is_2020} simulations: 1) the ``Cetus-New'' wrap, which overlaps the main part of the CPS at closer distances ($\approx$18 kpc compared to $\approx$30 kpc for the main wrap); and 2) the previously known C-20 stream \citep{ibata_charting_2021}. Meanwhile, \citetalias{thomas_cetus-palca_2022} used a sample of CPS stars they identified in SEGUE to guide an all-sky search for the CPS in \textit{Gaia} EDR3. After compiling a dense catalog of \textit{Gaia} member stars in the main wrap, they estimated the dwarf galaxy progenitor's stellar mass at $1.5\times10^6\,M_\odot$ based on the luminosity function of the stream members. 

In summary, the CPS is an example of a thick dwarf galaxy stream with a complex morphology. The main wrap, which we focus on in this work (owing to its proximity with Segue 2), is $\sim 100^\circ$ long, with a $1\sigma$ width and proper motion dispersion of $\sim 5^\circ$ and $\sim 0.2$ mas yr$^{-1}$ respectively (\citetalias{thomas_cetus-palca_2022}; \citealt{yuan_complexity_2022}). 

\subsection{The Segue 2 UFD}\label{subsec:seg2_info}

Segue 2 (discovered by \citealt{belokurov_discovery_2009}) is one of the faintest known MW satellites ($L_V = 900\,L_\odot$), possessing a stellar mass of just 1000 $M_\odot$ \citep{kirby_segue_2013}. As a UFD, Segue 2 is also extremely DM-dominated (mass-to-light ratio of $\approx360$ within its half-light radius of 46 pc; \citealt{kirby_segue_2013}), making it a very useful system for studying the nature of DM at small scales. 

Basic insights about Segue 2's DM halo can be gained from the kinematics of its stars, such as dynamical mass estimates \citep{belokurov_discovery_2009, kirby_segue_2013} and constraints on DM particle models, for instance fuzzy DM \citep{dalal_excluding_2022}. However, stellar kinematics can only probe the halo at scales similar to the half-light radius. The interaction of Segue 2 and the CPS presents a unique opportunity to constrain the mass and density profile of Segue 2's DM halo at kiloparsec scales via its gravitational influence on the CPS. 

There is ongoing debate about whether Segue 2 formed from the tidal stripping of an initially higher-mass dwarf or inside a very low-mass DM halo. The primary evidence that Segue 2 is a stripped remnant stems from its metallicity of [Fe/H] $\approx -2.2$, which places it well above the luminosity-metallicity relation extrapolated from brighter satellites \citep{kirby_segue_2013}. The tidal stripping scenario was also supported by \citet{belokurov_discovery_2009}, who found evidence for tidal tails around Segue 2, though \citet{kirby_segue_2013} could not confirm this with a larger spectroscopic sample of member stars. 

The evidence against the tidal stripping scenario comes from comparisons between Segue 2's tidal radius and its size. \citet{kirby_segue_2013} find that Segue 2 is too small to be undergoing tidal stripping at its present location. More recent studies that take Segue 2's orbit into account \citep{simon_faintest_2019, pace_proper_2022} have argued that Segue 2 is only at risk of stripping if its velocity dispersion is much lower than the upper limit of 2.2 km s$^{-1}$ measured by \citet{kirby_segue_2013}.

We note that regardless of whether Segue 2 has lost a significant fraction of its peak stellar mass to tidal stripping, its DM halo is likely more susceptible to stripping than its stars \citep[e.g.,][]{smith_preferential_2016}. In any case, its DM halo must once have been massive enough to retain enough supernova ejecta for Segue 2 to reach its present-day metallicity. \citet{kirby_segue_2013} used this line of reasoning to argue that Segue 2's halo must have had a peak mass between (1-3)$\times10^9\,M_\odot$.

More generally, studies also disagree on the minimum halo mass required to host a luminous galaxy at the present day. Many authors have used cosmological zoom-in simulations that include the effects of a UV background during reionization to study the star formation histories of UFDs. 
For instance, the simulations of \citet{shen_baryon_2014} produced no halos with masses below $10^9\,M_\odot$ containing stars at the present day, while the simulations of 
\citet{sawala_chosen_2016} found that only 10\% of halos less massive than $7\times10^8\,M_\odot$ host galaxies at the present. 
\citet{wheeler_sweating_2015} showed UFDs typically inhabit halos between ($0.5$ - $3)\times 10^9\,M_\odot$, in agreement with \citet{jeon_connecting_2017}, who found that the mass threshold for retaining gas through reionization is $3\times10^9\,M_\odot$ (see also \citealt{jeon_probing_2019, 
jeon_highly_2021,
jeon_role_2021} for extensions to this work).  \citet{munshi_quantifying_2021} found galaxies with stellar masses of $\sim 10^3\,M_\odot$ inhabiting $(0.3-15)\times10^{8}\,M_\odot$ halos in their simulations. While these works disagree on the exact halo mass threshold for galaxy formation (in part due to resolution differences; see e.g., \citealt{munshi_quantifying_2021}), constraints from cosmological simulations tend to settle around $10^9\,M_\odot$.

As an alternative to cosmological hydrodynamic simulations, \citet{bland-hawthorn_ultrafaint_2015} used idealized simulations of isolated galaxies to show halos as light as $10^7\,M_\odot$ can retain gas during a single supernova. Both \citet{jethwa_upper_2018} and \citet{nadler_milky_2020} assigned stellar masses to halos in cosmological simulations from a stellar mass - halo mass relation fit to MW satellites while accounting for observational biases. \citet{jethwa_upper_2018} estimated that the peak mass of the faintest MW satellite in their sample (Segue 1) is $2.4\times10^8\,M_\odot$, while \citet{nadler_milky_2020} found that the faintest known MW satellites likely have halo masses below $3.2\times10^8\,M_\odot$. \citet{kim_missing_2018} performed completeness corrections on Sloan Digital Sky Survey MW satellite counts and found that the occupation fraction of subhalos drops below 0.5 below $10^8\,M_\odot$.

A measurement of Segue 2's total mass and density profile from its imprint on the CPS would provide clarity on both the level of tidal stripping experienced by Segue 2 as well as a precious data point on the extreme faint end of the stellar mass - halo mass relation that could be used to test the minimum halo mass predictions outlined above. In this paper, we consider four mass models and two density profile shapes for Segue 2 (see Section \ref{subsec:perturbation}). The masses are equally spaced between (0.5 - $10)\times10^9\,M_\odot$, and we consider both Plummer profiles (which fall off quickly with radius, approximating a tidally truncated halo; \citealt{plummer_problem_1911}) and a Hernquist profile (which better approximates a pristine halo; \citealt{hernquist_analytical_1990}). This range of models is chosen to span the possibility that Segue 2 may have formed in a ``typical'' UFD halo of $\sim 10^9\,M_\odot$ or in an initially much more massive halo that has experienced tidal stripping. 

\section{Cetus-Palca Stream Data} \label{sec:obs}

In this section, we describe the construction of our CPS member catalog from various surveys, including \textit{Gaia} EDR3, H3, and SEGUE. To determine the orbit of the CPS, our goal is to build a high-purity sample of stars with at least five measured phase-space coordinates (sky position, proper motion, and distance). Where possible, we also include spectroscopic radial velocities. Section \ref{subsec:coords} lists and describes the coordinate systems we reference throughout this paper, while the datasets and selection of CPS members is discussed in Section \ref{subsec:selection}.

\subsection{Coordinate Systems} \label{subsec:coords}

Here, we describe the conventions we adopt for coordinate systems referred to in this paper. Equatorial coordinates are given in ICRS right ascension ($\alpha$) and declination ($\delta$). We also make use of a ``natural'' coordinate system for the CPS (longitude along the stream's orbital plane $\phi_1$, and latitude above/below the orbital plane $\phi_2$), closely based upon the \texttt{Cetus-Palca-T21} frame published in \texttt{galstreams}. Unlike the small-circle frame presented in \citetalias{thomas_cetus-palca_2022}, we choose a heliocentric great-circle frame in which $\phi_1$ increases in the stream's direction of motion. The frame's polar axis points at ($\alpha$, $\delta$) = (290.66303924$^\circ$, -12.52105422$^\circ$), and the origin is ($\alpha$, $\delta$) = (22.11454259$^\circ$, -6.50712051$^\circ$). Unless otherwise noted, proper motions are not corrected for the solar reflex motion, and proper motions in the $\alpha$ and $\phi_1$ directions include the cosine term. We adopt the \texttt{astropy} v4.0 \citep{astropy:2013, astropy:2018, astropy:2022} definition of Galactocentric Cartesian coordinates. However, note that we modify the default \texttt{astropy} Galactocentric frame to use the \citet{mcmillan_mass_2011} values for the local standard of rest ($v_{\rm{circ}}(8.29\, \rm{kpc}) = 239$ km s$^{-1}$) and the \citet{schonrich_local_2010} values for the Solar peculiar velocity ($(U,V,W) = (11.1, 12.24, 7.25)$ km s$^{-1}$). Transformations between coordinate systems are performed with \texttt{astropy} and \texttt{gala} \citep{price-whelan_gala_2017}.

\subsection{Member Selection} \label{subsec:selection}

\begin{figure*}
    \centering
    \includegraphics[width=\textwidth]{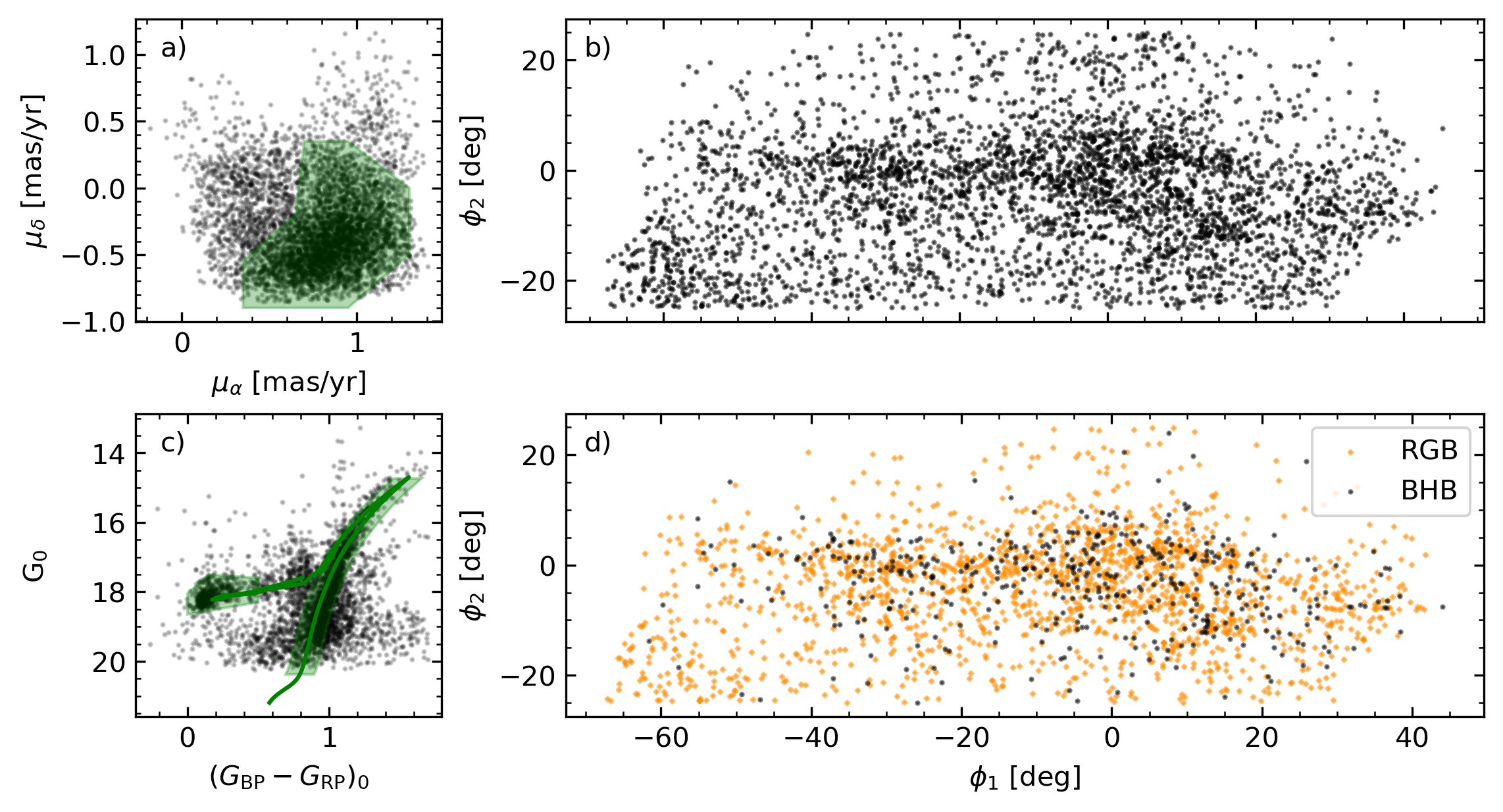}
    \caption{Initial selection of CPS members from the catalog of \citetalias{thomas_cetus-palca_2022}. \textit{a)} ICRS proper motions of 5336 stars with $P_{\rm{Cetus}} > 0.5$ and $|\phi_2| \leq 25^{\circ}$. Stars belonging to the stream form an overdensity in proper motion space, which we select with the shaded polygon. \textit{b)} On-sky positions of 3921 stars within the proper motion selection polygon in panel \textit{a} in the CPS's natural coordinates (see Section \ref{subsec:coords}). \textit{c)} Color-magnitude diagram of proper motion - selected stars. We use a 14 Gyr isochrone with [Fe/H] $= -1.93$ and shifted by a distance modulus of 17.7 (green line) to select BHB and RGB stars within the shaded regions. \textit{d)} On-sky positions of 2210 stars passing both the proper motion and color-magnitude diagram selections. BHB (RGB) stars are shown as black circles (orange diamonds).}
    \label{fig:sel}
\end{figure*}

\textbf{\citetalias{thomas_cetus-palca_2022} Catalog:} We begin with the CPS member catalog selected from \textit{Gaia} \citep{gaia_collaboration_gaia_2016} EDR3 \citep{gaia_collaboration_gaia_2021} as described in Section 4.3 of \citetalias{thomas_cetus-palca_2022}. For the full details of their selection, we refer the interested reader to \citetalias{thomas_cetus-palca_2022}, but provide an abridged version here. The sample is selected on the basis of color ($-0.5 \leq (G_{\rm{BP}} - G_{\rm{RP}})_0 \leq 1.7$), parallax ($-0.4 < \varpi < 0.2$ mas), proximity to the CPS orbital plane, and proper motion. Note that their orbital plane and proper motion selections are based on a different coordinate system for the CPS than we adopt in this paper (see Section \ref{subsec:coords}). 

In addition to these criteria, \citetalias{thomas_cetus-palca_2022} developed a model (their Equation 4) for the probability $P_{\rm{Cetus}}$ that a star in the sample belongs to the CPS (as opposed to the foreground or background), based on the stars' sky positions and proper motions. The bottom panel of their Figure 12 shows the sky positions of the stars in their catalog with $P_{\rm{Cetus}} > 0.2$. 

We note that \citetalias{thomas_cetus-palca_2022} performed this selection to maximize completeness for the purposes of estimating the CPS progenitor's mass. However, a kinematic study of the stream such as the one presented here requires that we apply additional criteria to maximize purity. To begin, we remove stars with $P_{\rm{Cetus}} < 0.5$ and $|\phi_2| > 25^{\circ}$. 

\textbf{Proper Motion Selection:} Figure \ref{fig:sel} illustrates the next steps of the selection process. Panel a shows the proper motions of potential stream stars. An overdensity in proper-motion space is indicated by the shaded polygon, which we fit by eye to maximize the contrast between the stream and the background. Panel b shows the stars encompassed by the polygon in CPS coordinates. Already, the stream is apparent against the remaining contamination. 

\textbf{Color-Magnitude Diagram Selection:} To further refine our selection, we use the color-magnitude diagram in panel c, which contains all of the proper motion - selected stars. \citetalias{thomas_cetus-palca_2022} found that the CPS is well-described by a 14 Gyr-old MIST \citep{choi_mesa_2016, dotter_mesa_2016} isochrone with [Fe/H] = -1.93 and a distance modulus of 17.7. We identify the BHB and red giant branch (RGB), and select stars belonging to both evolutionary stages within the shaded areas associated with this isochrone. In panel d, we are left with a relatively pure sample of CPS stars when compared with the original \citetalias{thomas_cetus-palca_2022} catalog, though some contamination remains. 

\begin{figure*}
    \centering
    \includegraphics[width=\textwidth]{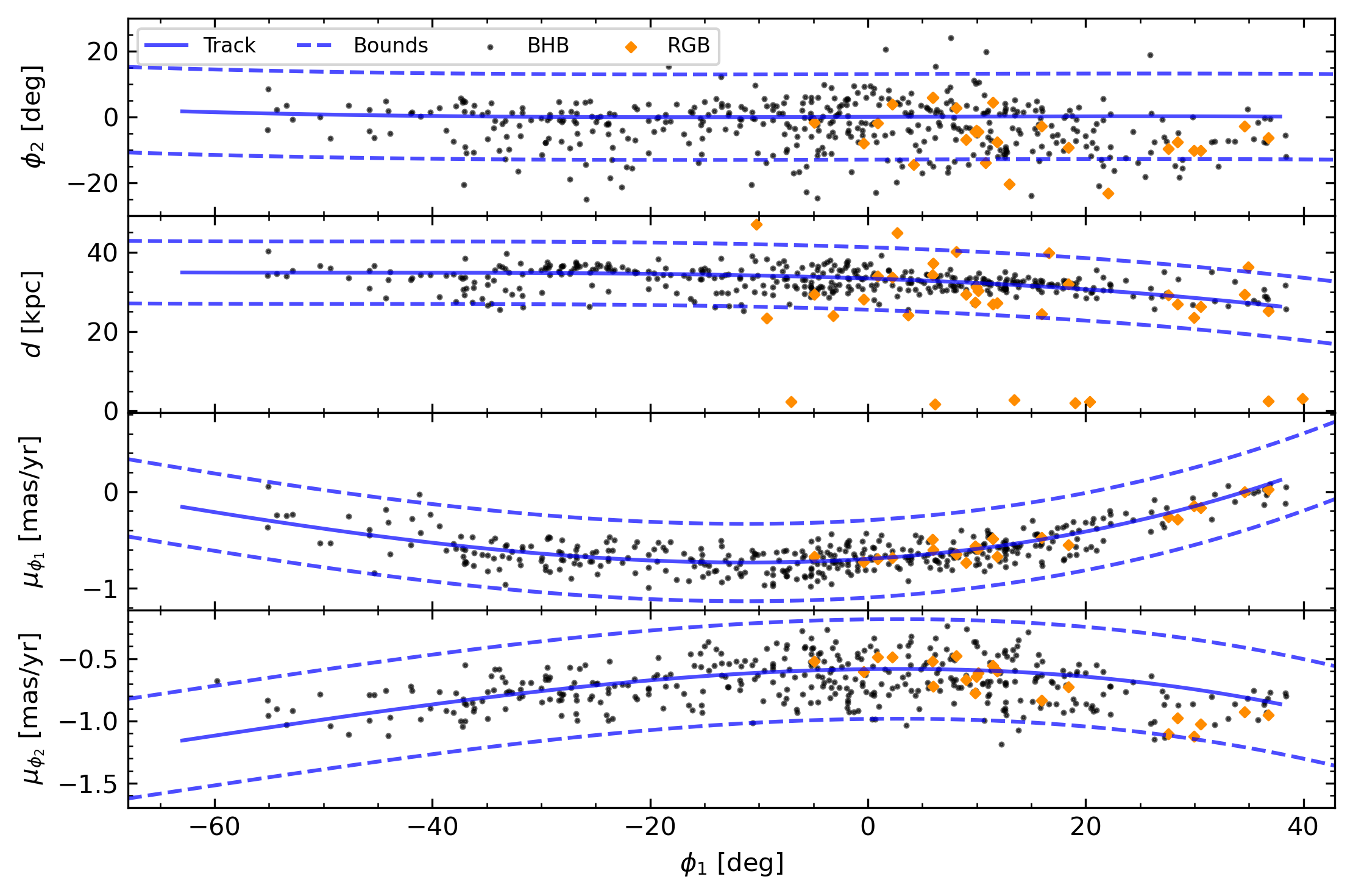}
    \caption{Track selection of 5-D CPS members. Each panel shows one phase-space coordinate ($\phi_2$, distance, proper motions) as a function of $\phi_1$. Black circles denote BHB stars selected in Figure \ref{fig:sel}, and orange diamonds show RGB stars selected in Figure \ref{fig:sel} and cross-matched with H3. The solid lines show the \texttt{Cetus-Palca-T21} track from \texttt{galstreams} \citep{mateu_galstreams_2023}, and the dashed lines in each panel show the spread about the track used to select the stream as described in the text. Inspired by \citet{koposov_piercing_2019}, \citet{li_broken_2021}, and  \citet{koposov_s5_2023}, each panel contains stars selected from the other panels but not itself, i.e. stars in the $\phi_2$ panel are selected with distance and both proper motions, but not $\phi_2$. The final selection of stream members includes only stars that pass the selection in all four panels.}
    \label{fig:track_sel}
\end{figure*}

\textbf{Radial Velocities and Distances:} Thus far, our sample utilizes on sky positions and proper motions from \textit{Gaia}. We now need distances and radial velocities for the stars in this sample to complete the full 6-dimensional dataset.
To this end, we cross-match the remaining stars with the H3 survey \citep{conroy_mapping_2019}, identifying 47 matches from our RGB sample, and 8 matches from our BHB sample. 

Performed with the MMT's Hectochelle spectrograph \citep{szentgyorgyi_hectochelle_2011}, H3 consists of over 300,000 $\rm{R}=32,000$ spectra of MW halo stars selected from \textit{Gaia} parallaxes and Pan-STARRS \citep{chambers_pan-starrs1_2016} $r$-band magnitudes. H3 uses the \texttt{MINESweeper} pipeline \citep{cargile_minesweeper_2020} to derive stellar parameters such as metallicities, $\alpha$-abundances, radial velocities, and spectrophotometric distances, and we use the latter two properties to fill in the remaining dimensions for our cross-matched CPS stars. 

\citetalias{thomas_cetus-palca_2022} also identified 91 SEGUE \citep{yanny_segue_2009} stars with spectroscopic radial velocities as likely CPS members, and we include a subset of 22 of these stars that meet the other selection criteria imposed so far. Distances for these SEGUE stars are provided by the machine-learning model presented in \citetalias{thomas_cetus-palca_2022}. One star is duplicated in our SEGUE and H3 data, and we take the H3 values for its distance and radial velocity, though both measurements are consistent between the surveys.

Next, to obtain distances for the BHB stars that were not matched with H3 or SEGUE, we make use of the \citet{deason_milky_2011} relation between the color and absolute magnitude of BHB stars, re-calibrated for the \textit{Gaia} EDR3 photometric system by \citetalias{thomas_cetus-palca_2022}:

\begin{eqnarray} \label{eqn:bhb}
    M_{\rm{G,BHB}} = && -0.266 (G_{\rm{BP}} - G_{\rm{RP}})_0^3 + 2.335 (G_{\rm{BP}} -  G_{\rm{RP}})_0^2 \nonumber \\ 
    && - 1.955 (G_{\rm{BP}} - G_{\rm{RP}})_0 + 0.794
\end{eqnarray}

Using this relation, we derive photometric distances to every BHB star in our remaining sample. Relative distance errors are assumed to be 5\% (\citealt{deason_milky_2011}; \citetalias{thomas_cetus-palca_2022}). As no corresponding relation exists for RGB stars, we keep only the RGB stars that have distances measured from H3 (or from \citetalias{thomas_cetus-palca_2022}'s model for the SEGUE stars).

\textbf{Stream Track Selection:} In Figure \ref{fig:track_sel}, we plot the BHB stars and RGB stars with known distances alongside the \texttt{Cetus-Palca-T21} CPS track from \texttt{galstreams}. As a last step to increase the purity of our sample of 5-D stars, we remove stars far from this known stream track, similar to the method used by \citet{koposov_s5_2023}. The dashed blue lines show the region used to select stream stars in each dimension. We keep stars within $13^{\circ}$ of the $\phi_2$ track, 7.87 kpc of the distance track (the same width as the $\phi_2$ selection assuming a distance of 35 kpc), and 0.4 mas yr$^{-1}$ of the proper motion tracks.
These bounds are designed to fully encompass the stream given recent estimates of its 1$\sigma$ width ($\sim$ $5^\circ$, $\sim$ $0.2$ mas yr$^{-1}$; \citetalias{thomas_cetus-palca_2022}; \citealt{yuan_complexity_2022}) while discarding contaminants that are far from the stream track in position and/or velocity. 

As in Figure 2 of \citet{koposov_piercing_2019}, Figure 2 of \citet{li_broken_2021}, and Figure 1 of \citet{koposov_s5_2023}, each panel of Figure \ref{fig:track_sel} shows the stars selected using constraints placed on the other three dimensions. For example, the stars shown in the top panel ($\phi_2$ versus $\phi_1$) are selected using the distance and proper motion cuts but \textit{not} $\phi_2$. This allows us to confirm that our sample is relatively pure, i.e. no more than $11\%$ of stars in any given dimension lie outside our chosen bounds while passing the selections made in the other dimensions. Ultimately, we keep the stars that lie within the chosen bounds of the track in all four dimensions.

\subsection{Radial Velocity Track} \label{sec:rv}

\begin{figure}
    \centering
    \includegraphics[width=\columnwidth,trim=5mm 5mm 0mm 0mm]{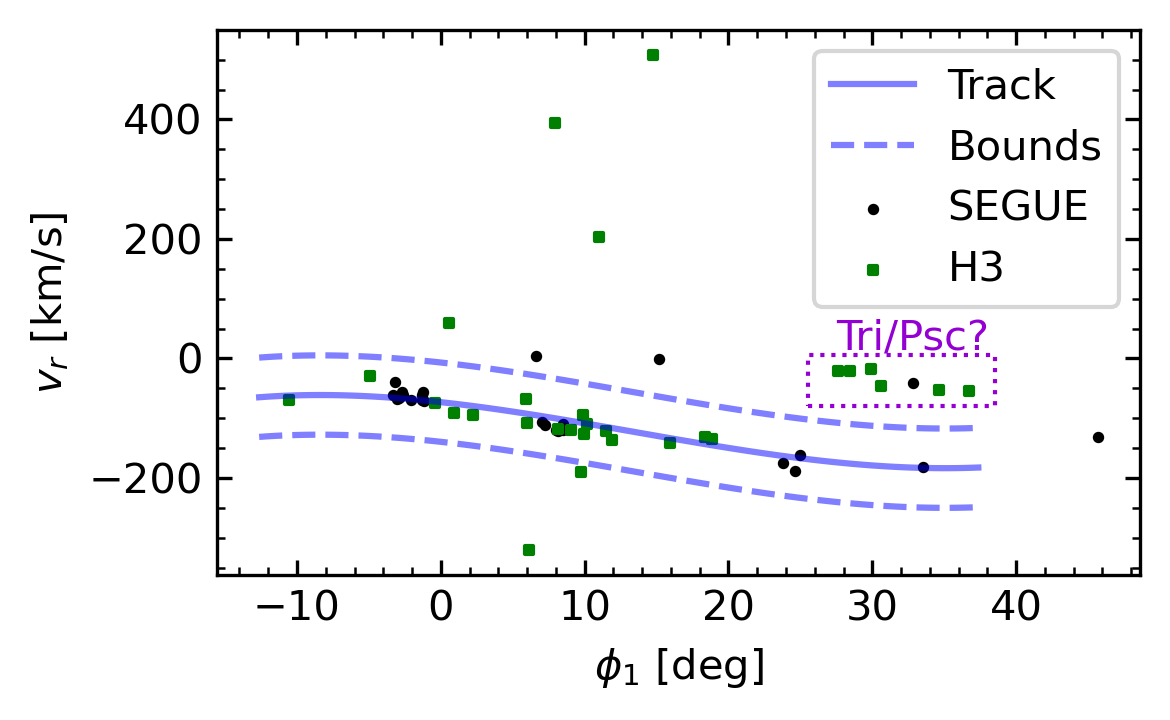}
    \caption{$v_r$ vs. $\phi_1$ for 55 stars with measured radial velocities after the selection process described in Section \ref{subsec:selection}. SEGUE (H3) stars are shown as black circles (green squares). The  CPS is clearly visible in radial velocity space between $-12^\circ \lesssim \phi_1 \lesssim 38^{\circ}$. In addition to the CPS, we identify a group of stars around $\phi_1 \sim 33^\circ$ and $v_r \sim -20$ km s$^{-1}$ that have radial velocities consistent with the Tri/Psc stream (purple box; see Section \ref{subsec:tripsc}), as well as some contamination from stars clearly not associated with either stream. As described in the text, we fit a radial velocity track to the CPS stars (solid blue line), and keep the 39 stars within 66.4 km s$^{-1}$ of this track (dashed blue lines) as members of our 6-D sample. 
    }
    \label{fig:rv}
\end{figure}

With our catalog of CPS members in-hand, we now turn to the subset of these members with measured radial velocities to determine the radial velocity track of the CPS. Figure \ref{fig:rv} shows the heliocentric radial velocities ($v_r$) of the stars for which we have radial velocity measurements. While the CPS is clearly visible in radial velocity space, we see another group of stars near $\phi_1 \sim 33^\circ$ and $v_r \sim -20$ km s$^{-1}$ (inside the purple box). These stars have radial velocities consistent with the Triangulum / Pisces stream (Tri/Psc; \citealt{bonaca_cold_2012, martin_kinematic_2013, martin_pandas_2014}), which has a tentative association with the CPS (\citealt{bonaca_orbital_2021}; \citetalias{thomas_cetus-palca_2022}; \citealt{yuan_complexity_2022}). We discuss these stars further in Section \ref{subsec:tripsc}.

After removing the candidate Tri/Psc stars, we are left with CPS stars and a small number of contaminants. To find the radial velocity track of the CPS, we calculate the median $v_r$ in $10^\circ$ wide $\phi_1$ bins, shifted by $5^\circ$ along the stream, and then fit a third-degree polynomial to the binned medians. The result is shown as the solid blue line in Figure \ref{fig:rv}. To remove the remaining contaminants, we use the same method as in Figure \ref{fig:track_sel}, keeping stars with radial velocities less than 66.4 km s$^{-1}$ from the track (dashed blue lines). This velocity limit of 66.4 km s$^{-1}$ corresponds to 0.4 mas yr$^{-1}$ in proper motion space assuming a distance of 35 kpc, consistent with the proper motion cuts in Figure \ref{fig:track_sel}. 

Ultimately, we are left with a catalog containing two samples of CPS stars:
\begin{itemize}
    \item \textbf{5-D Sample:} 364 stars with five phase-space dimensions measured: sky positions, proper motions, and distances. 
    \item \textbf{6-D Sample:} 39 stars with all six phase-space dimensions measured via the addition of spectroscopic radial velocities. 
\end{itemize}

\noindent These stars span $\sim 90^\circ$ on the sky at roughly constant $\alpha \sim 30^\circ$ between $-60^\circ \lesssim \delta \lesssim 30^\circ$. The stream shows a distance gradient of 7 kpc, from $\sim35$ kpc at the trailing edge to $\sim28$ kpc at the leading edge, in agreement with the pure sample of BHB stars identified by \citetalias{thomas_cetus-palca_2022}. These stream properties are also consistent with the previous detections of this wrap by \citet{yuan_revealing_2019} and \citet{yuan_complexity_2022}. Note that we do not search for the diffuse component of the stream found in the Galactic north by \citet{yuan_revealing_2019} and \citet{yuan_complexity_2022}, nor for the additional wraps identified by \citet{yuan_complexity_2022}, as these are far from Segue 2 and are not expected to be recently affected by its passage. 

\section{Stream Orbit Fitting} \label{sec:orbit}

In this section we estimate the orbit of the CPS based on the 5- and 6-D member catalogs assembled in the previous section. This allows us to track the location of the CPS stars backwards in time and identify the location and timing of closest approach between Segue 2 and the CPS. In Section \ref{subsec:gals}, we describe our model of a combined MW / LMC gravitational potential. As no remnant of the CPS progenitor has been conclusively identified (e.g., \citealt{yuan_revealing_2019, yuan_complexity_2022, chang_is_2020}; \citetalias{thomas_cetus-palca_2022}), we utilize two complementary methods for inferring the stream's orbit from the kinematics of its stars. In Section \ref{subsec:fits}, we fit the orbit of the CPS based on a maximum-likelihood method used by \citet{price-whelan_off_2018} and \citet{bonaca_stellar_2025}, and in Section \ref{subsec:repstar}, we use the method of \citet{chang_is_2020} to find a stellar tracer that is exemplary of the stream. 
We refer to the orbits inferred with these methods as the ``Fit Orbit'' and ``Tracer Orbit,'' respectively. Using two different orbit modeling approaches captures the uncertainties involved with modeling the stream while allowing us to study two different CPS - Segue 2 impact geometries in later sections.

\subsection{Galaxy Models} \label{subsec:gals}

\begin{table}[]
    \caption{
    Galaxy model parameters.
    } \label{tab:gal_models}
    \centering
    \begin{tabular}{c c c}
        \hline
        \hline
         Parameter & Value & Unit \\
         \hline
         \multicolumn{3}{c}{Milky Way} \\
         \hline
         $M_\mathrm{vir}$  & $1\times10^{12}$ & $M_\odot$\\
         $R_\mathrm{vir}$  & 262.76 & kpc\\
         $c_\mathrm{vir}$  & 9.86 & -\\
         $M_\mathrm{disk}$ & $6.5 \times 10^{10}$ & $M_\odot$ \\
         $R_\mathrm{disk}$ & 3.5  & kpc\\
         $z_\mathrm{disk}$ & 0.53  & kpc\\
         $M_\mathrm{H,bulge}$ & $1\times10^{10}$ & $M_\odot$\\
         $a_\mathrm{bulge}$ & 0.7 & kpc\\
         \hline
         \multicolumn{3}{c}{LMC} \\
         \hline
         $M_\mathrm{H}$ & $1.8\times10^{11}$  & $M_\odot$\\
         $a$ & 20 & kpc\\
         \hline
         \hline
    \end{tabular}
    \tablenotetext{}{NOTE - To calculate virial parameters for the MW, we use the \citet{planck_collaboration_planck_2020} cosmology. From top to bottom, this table includes, for the MW: halo virial mass as defined by \citet{bryan_statistical_1998}, i.e. the radius where the average density of the halo is $\Delta_{\rm{vir}} =327.4$ times the average matter density of the universe ($\Omega_m \rho_\mathrm{crit}$); halo virial radius from Equation A1 of \citet{van_der_marel_m31_2012}; halo concentration; disk mass; disk radial scale; disk vertical scale; bulge mass; and bulge scale. For the LMC: total halo mass; halo scale.}
\end{table}

To find the orbit of the CPS, we require an assumption of the underlying gravitational potential. The MW model we adopt is described in \citet{patel_orbits_2017} and \citet{patel_orbital_2020} (their MW1). It is composed of an analytic NFW halo \citep{navarro_structure_1996}, a Miyamoto-Nagai disk \citep{miyamoto_three-dimensional_1975}, and a Hernquist bulge \citep{hernquist_analytical_1990}, which remain fixed with respect to each other during integration. We also include the LMC, owing to its substantial influence on the orbits of MW halo tracers \citep[e.g., review by][and references therein]{vasiliev_effect_2023}, such as stellar streams \citep[e.g.,][]{vera-ciro_constraints_2013, gomez_and_2015, erkal_modelling_2018, erkal_total_2019, shipp_proper_2019, shipp_measuring_2021, ji_kinematics_2021,  vasiliev_tango_2021, koposov_s5_2023, lilleengen_effect_2023}. Our LMC model is described in \citet{garavito-camargo_hunting_2019} (their LMC3), and consists of an analytic Hernquist halo. The parameters of these galaxy models are summarized in Table \ref{tab:gal_models}. 
 
The LMC's primary effect on stellar streams stems from the reflex motion of the MW about the combined MW+LMC barycenter \citep[e.g.,][]{weinberg_production_1995, gomez_and_2015, erkal_total_2019, petersen_reflex_2020, ji_kinematics_2021, vasiliev_tango_2021}. We account for this by allowing the centers of mass of the MW and LMC to move in response to each other. However, as the potentials are analytic and therefore remain rigid and symmetric, we do not take into account the shape distortions of the MW and LMC halos during the LMC's infall \citep{erkal_total_2019, garavito-camargo_hunting_2019, cunningham_quantifying_2020, erkal_equilibrium_2020, petersen_reflex_2020, garavito-camargo_quantifying_2021, makarov_lmc_2023, vasiliev_dear_2024, chandra_all-sky_2024, sheng_uncovering_2024, yaaqib_radial_2024}. The consequences of this choice are discussed further in Section \ref{ssubsec:rigid_pots}. We use the values given in \citet{kallivayalil_third-epoch_2013} for the LMC's present-day Galactocentric position and velocity vector.

\subsection{Likelihood Model} \label{subsec:fits}

To fit the orbit of the CPS, we construct a maximum-likelihood-based model similar to the one used by \citet{price-whelan_off_2018} and \citet{bonaca_stellar_2025}, which searches for the orbit of a test particle that best traces the track of the stream. The \texttt{Cetus-Palca-T21} track from \texttt{galstreams} is missing radial velocity information and is not informed by orbit fitting \citep{mateu_galstreams_2023}. In contrast, a fit orbit will describe the stream more completely and will also be consistent with the potentials described in the previous section. 

We note that, especially near apocenter, the track of a stream does not necessarily trace 
the stream
orbit accurately, 
which can lead to additional systematic errors in fitting the MW potential if this assumption is made \citep{eyre_locating_2009, eyre_mechanics_2011, sanders_streamorbit_2013, sanders_streamorbit_2013-1}. However, our aim is not to fit the potential of the MW / LMC system; rather we seek a description of the stream's location as a function of time within a few hundred Myr of the present. Additionally, \citet{chang_is_2020} found that the CPS traces the orbit of its progenitor over the past $\sim$ two wraps. Therefore, finding the orbit that best traces the stream track is sufficient for our purposes. 

To ensure the orbit covers the $\phi_1$ range of the stream, we place a test particle ahead of the stream and integrate its orbit backwards in time for long enough that it passes behind the last stream star. In detail, we initialize the test particle at a stream longitude of $\phi_{1,0} = 45^\circ$, with the other phase-space coordinates given as model parameters: 

\begin{equation}
    \vec{P} = [\phi_{2,0},\: d_0,\: \mu_{\alpha,0},\: \mu_{\delta,0},\: v_{r,0}].
\end{equation}

\noindent Using \texttt{gala}, we then integrate the test particle's orbit backwards in time for 500 Myr in our combined MW+LMC potential. This test particle orbit can be thought of as a  discrete function which gives the ``dependent'' phase-space coordinates $(\phi_{2},\: d,\: \mu_{\alpha},\: \mu_{\delta},\: v_{r})$ as a function of $\phi_1$. For each dependent coordinate, we interpolate the discrete timesteps of the orbit with a cubic spline, obtaining a continuous function describing the test particle orbit as a function of $\phi_1$ and $\mathbf{P}$:

\begin{eqnarray}\label{eqn:model}
    \vec{\mathcal{O}}(\phi_1, \vec{P}) = &&[\phi_{2}^{orbit}(\phi_1),\: d^{orbit}(\phi_1),\: \mu_{\alpha}^{orbit}(\phi_1),\: \nonumber \\
    &&\mu_{\delta}^{orbit}(\phi_1),\: v_{r}^{orbit}(\phi_1)]. 
\end{eqnarray}

\noindent We calculate the log-likelihood that the test particle orbit passes through the stream based on the measured locations of the CPS stars selected in Section \ref{sec:obs}:

\begin{equation} \label{eqn:likelihood}
    \mathrm{ln} (\mathcal{L}) = \sum_{i} \sum_{j} \left( -\frac{1}{2}\frac{(\mathcal{O}_i (\phi_{1,j}, \vec{P}) - x_{ij})^2}{\delta x_{ij}^2} - \frac{1}{2} \mathrm{ln} (2\pi\,\delta x_{ij}^2) \right),
\end{equation}

\noindent where the subscript $i$ labels the five dependent coordinates such that the position of the $j$th star is ($\phi_{1,j}$, $x_{ij}$), and $\delta x_{ij}$ similarly denotes the uncertainty on the $j$th star's position in the $i$th coordinate. The uncertainties we use for this calculation are as follows.

For the $\phi_2$ uncertainties, the width of the stream in $\phi_2$ is on the order of degrees, \textit{much} larger than the \textit{Gaia} astrometry errors which are typically less than a milliarcsecond. This makes the observational errors a poor estimate of the uncertainty in the orbit given the width of the stream. Instead, we measure the $1\sigma$ width of the stream in $\phi_2$ (denoted $\sigma_{\phi_2}$) and set $\delta \phi_2$ equal to this for every star. To calculate $\sigma_{\phi_2}$, we first find the standard deviation of the star positions in $\phi_2$ in a $10^\circ$-wide $\phi_1$ window shifted by $5^\circ$ along the stream. After verifying the width of the stream is approximately constant with $\phi_1$ (see the solid line in the top row of panels in Figure \ref{fig:mockstreams}), we set $\sigma_{\phi_2}$ equal to the mean of the binned widths, which is $4.8^\circ$. This width is consistent with other recent estimates of the CPS's width (\citetalias{thomas_cetus-palca_2022}; \citealt{yuan_complexity_2022}).

The distance uncertainties are those reported by SEGUE or H3 where available, or assumed to be 5\% for the BHB photometric distances (\citealt{deason_milky_2011}; \citetalias{thomas_cetus-palca_2022}), as discussed in Section \ref{subsec:selection}. 
We keep the proper motions in equatorial coordinates (as opposed to CPS coordinates) so that we can use the reported errors from \textit{Gaia} without needing to transform these to the CPS frame. For stars with measured radial velocities, we use the uncertainties reported by SEGUE or H3. For stars without measured radial velocities, we use a value of 0 km s$^{-1}$ and an uncertainty of 1000 km s$^{-1}$ to ensure that these stars do not meaningfully contribute to the likelihood in the $v_r$ dimension.

To find the maximum of the likelihood, we use the BFGS algorithm available in \texttt{scipy.optimize} \citep{virtanen_scipy_2020}. 

\subsection{Exemplary Tracer Star} \label{subsec:repstar}

\begin{figure}
    \centering
    \includegraphics[width=\columnwidth, trim=5mm 5mm 4mm 0mm]{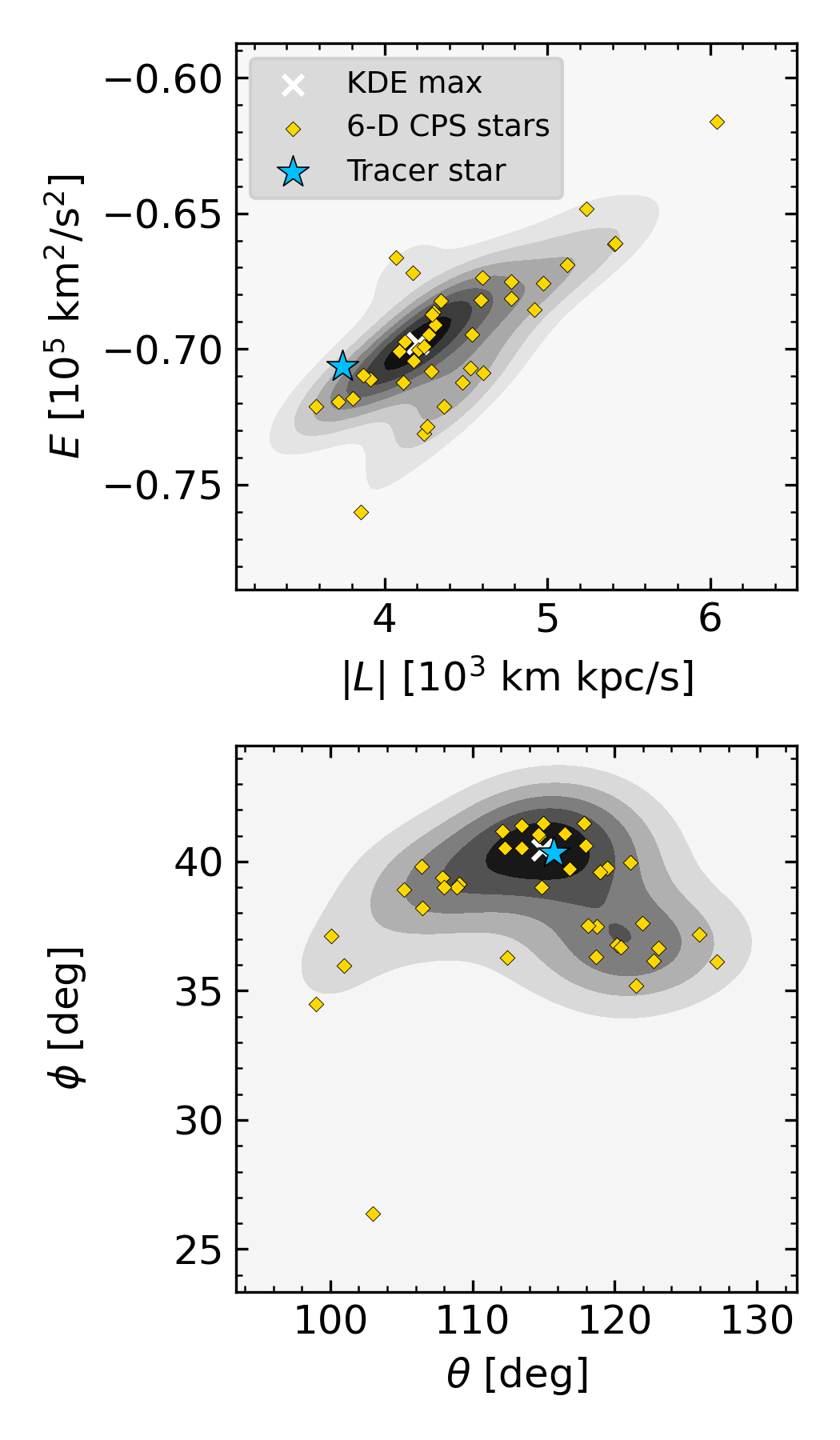}
    \caption{Selection of an example 6-D tracer star for the CPS. The distribution of 6-D stars in energy-angular momentum (top panel) and orbital pole (bottom panel) spaces is shown by the yellow diamonds and as contours of a kernel density estimate (KDE). The max of the KDE in each panel is shown with a white cross. The ``exemplary'' star selected as the stream tracer
    (blue star) is close to the max of the orbital pole distribution and below the max of the $(E,|\vec{L}|)$ distribution (c.f. Figure 3 of \citet{chang_is_2020}).}
    \label{fig:repstar}
\end{figure}

For a complementary method of estimating the orbit of the CPS, we use a slight modification of the method presented in \citet{chang_is_2020} to identify a single tracer star whose orbit is exemplary of the rest of the stream. For each star in our 6-D sample, we calculate the orbital energy $E$ and orbital angular momentum $\vec{L}$ about the MW's center of mass in Galactocentric coordinates. From the angular momentum, we also calculate the stars' orbital poles ($\theta$, $\phi$), where $\theta = \rm{arccos}(L_z/|\vec{L}|)$ is the polar angle from the Galactocentric $z$-axis, and $\phi = \rm{arctan2}(L_x, |\vec{L}|)$ is the azimuthal angle from the Galactocentric $x$-axis.\footnote{We note that \citet{chang_is_2020} use a different coordinate system to report the angular momentum vector and orbital poles of their CPS stars, so care must be taken when directly comparing our Figure \ref{fig:repstar} to their Figure 3.} 

Similar to Figure 3 of \citet{chang_is_2020}, Figure \ref{fig:repstar} shows the distribution of the 6-D stars in the ($E$, $|\vec{L}|$) and ($\theta$, $\phi$) spaces in the top and bottom panels, respectively. For each space, we construct a kernel density estimate (KDE) of the distribution, shown by the shaded contours. To choose an exemplary tracer star, we sum the values of the two KDEs (normalized such their maxima are equal to 1) at the position of each star and choose the star with the largest combined value. The chosen tracer (blue star) is very close the maximum of the orbital pole distribution (white cross) and below the maximum of the ($E$, $|\vec{L}|$) distribution. We measure this star's closest approach with Segue 2 by the procedure described in Section \ref{subsec:perturbation}, finding that the star's minimum separation with Segue 2 over the past 500 Myr is 9.11 kpc.

\subsection{Orbits}\label{subsec:orbit_results}

\begin{figure*}
    \centering
    \includegraphics[width=\textwidth]{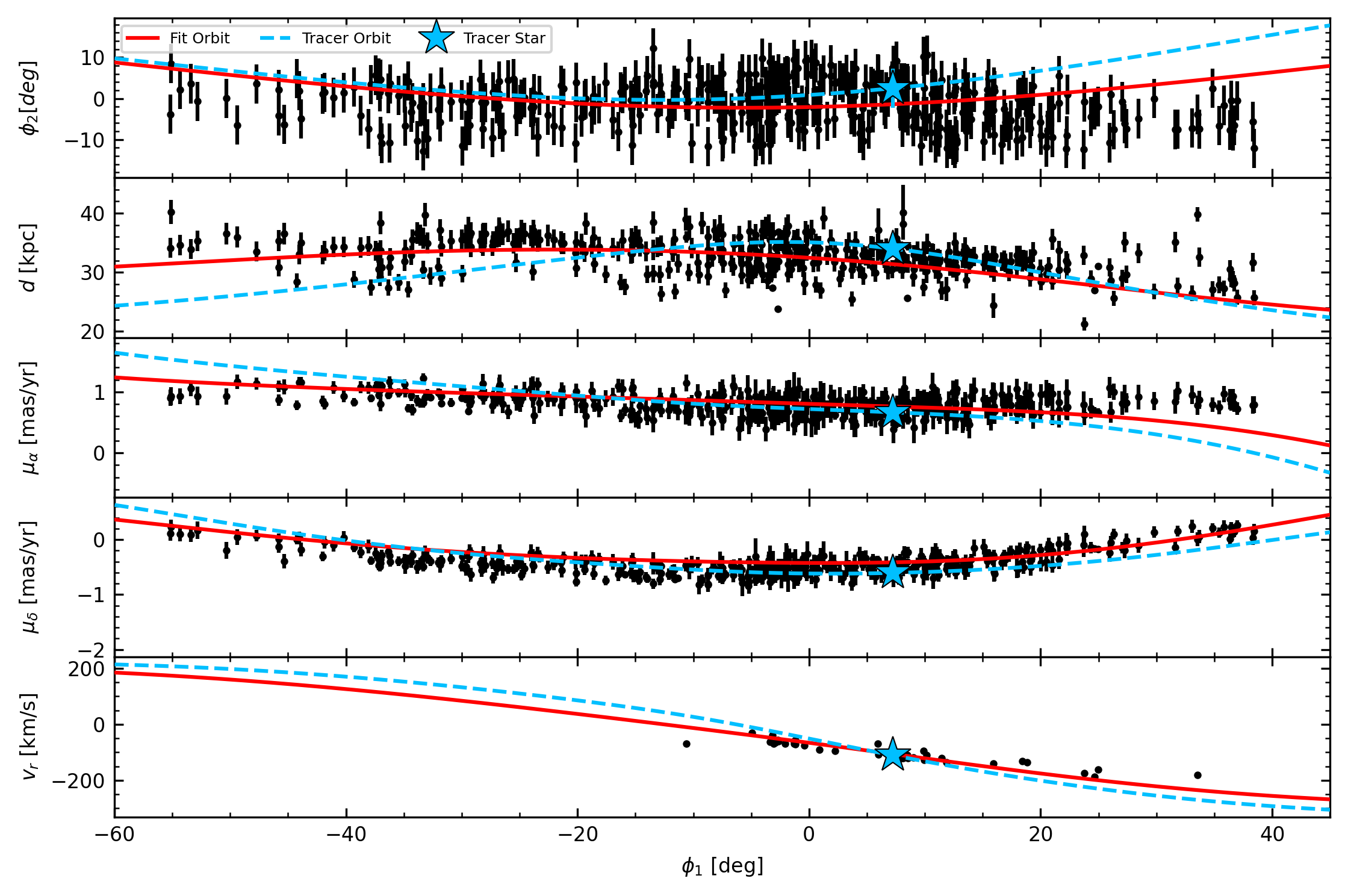}
    \caption{Model orbits for the CPS. Each panel shows the positions of the stars and their associated errors in the phase-space coordinates used to fit the CPS orbit as described in Section \ref{subsec:fits}, plotted against $\phi_1$. The position of the exemplary tracer star found in Section \ref{subsec:repstar} is shown with a blue star. The top four panels show all of the stars in our CPS catalog, while the $v_r$ panel (bottom) shows only the 6-D sample for clarity. Note the radial velocity error bars tend to be smaller than the points. The orbit of the test particle with initial conditions that maximize Equation \ref{eqn:likelihood} is shown as the solid red line (``Fit Orbit"), while the orbit of the tracer star is shown as the dashed blue line (``Tracer Orbit"). The Fit Orbit traces the stream's location over nearly its entire length. The Tracer Orbit reproduces the overall behavior of the stream but fits less well than the Fit Orbit.} 
    \label{fig:orbits}
\end{figure*}

\begin{table}[]
    \caption{
    Initial Conditions for CPS Orbits.
    } \label{tab:ics}
    \centering
    \begin{tabular}{c c c c}
        \hline
        \hline
         Parameter & Fit Orbit & Tracer Orbit & Unit\\
         \hline
         $x$   & -16.39 & -19.37 & kpc\\
         $y$   & 19.42 & 11.27 & kpc\\
         $z$   & -10.78 & -30.08 & kpc\\
         $v_x$ & 73.43 & 7.24 & km s$^{-1}$\\
         $v_y$ & 41.82 & 79.61 & km s$^{-1}$\\
         $v_z$ & 171.77 & 69.56 & km s$^{-1}$\\
         \hline
         \hline
    \end{tabular}
    \tablenotetext{}{NOTE - Initial Galactocentric positions and velocities for the test particles describing the Fit and Tracer CPS Orbits.}
\end{table}

\begin{figure}
    \centering
    \includegraphics[width=\columnwidth, trim=5mm 5mm 0mm 0mm]{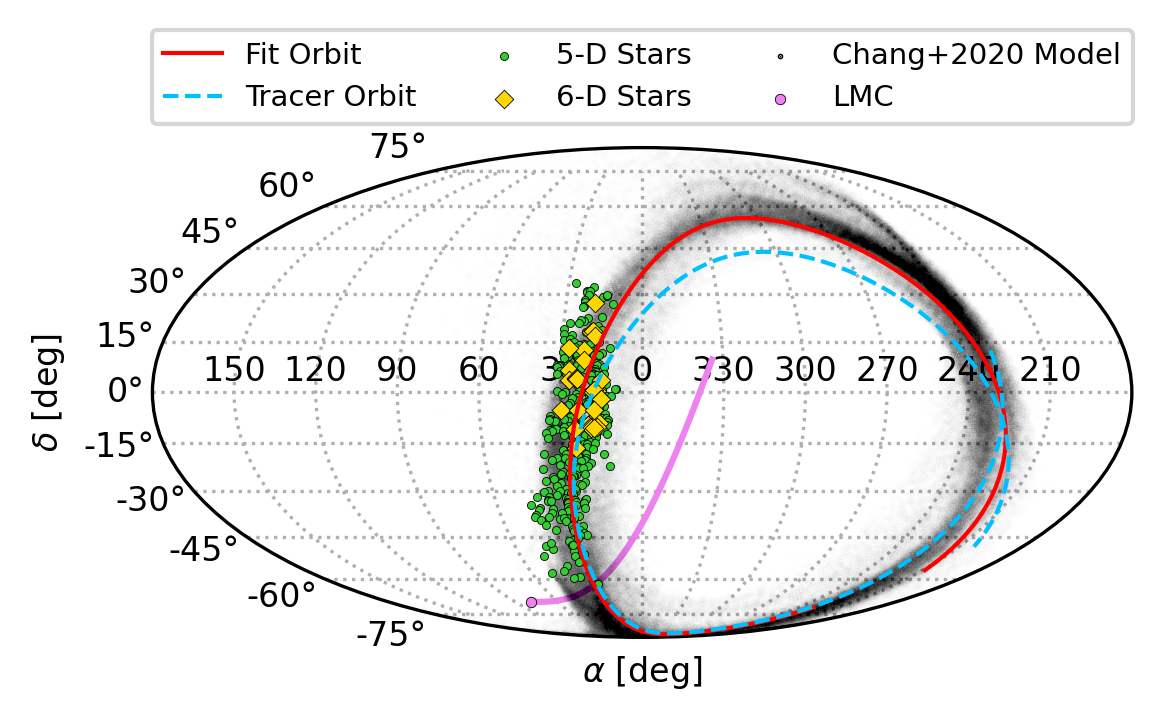}
    \caption{Sky positions in Mollweide projection of our CPS stars and orbit models compared to the CPS \textit{N}-body model presented in \citet{chang_is_2020}. Our CPS stars are split into the 5-D sample (green circles) and 6-D sample (yellow diamonds). The Fit and Tracer Orbits are shown as the solid red and dashed blue lines, respectively. The LMC and its past orbit (over the last 500 Myr) are shown as the purple point and line, though note that the LMC's orbit lies at larger distances than the CPS. The CPS model from \citet{chang_is_2020} is shown as the black points in the background. Our CPS member stars and orbits agree very well with the \citet{chang_is_2020} model, though the Tracer Orbit deviates slightly from the model in the North.}
    \label{fig:sky_orbit}
\end{figure}

The CPS orbits derived from the fitting procedure in Section \ref{subsec:fits} and the tracer star selection procedure in Section \ref{subsec:repstar} are presented in Figure \ref{fig:orbits}. Note that the proper motions are given in equatorial coordinates while the other dimensions are in CPS coordinates, as in Equation \ref{eqn:likelihood}. Both orbits are integrated 500 Myr forwards and 500 Myr backwards from their initial conditions (summarized in Table \ref{tab:ics}). For the ``Fit CPS Orbit,'' (solid red line) the initial conditions are those which maximize Equation \ref{eqn:likelihood}, while the ``Tracer CPS Orbit'' (dashed blue line) is integrated from the present-day position and velocity of the tracer star. 
Note that the bottom panel ($v_r$ vs. $\phi_1$) shows only stars with measured radial velocities for clarity. In this panel, the radial velocity error bars are typically smaller than the points. 

The Fit Orbit traces the location of the stream well, lying within the width of the stream over its entire length in every dimension except $\mu_\alpha$, in which the orbit diverges from the stream slightly at $\phi_1 > 25^\circ$. The best agreement between the stream and Fit Orbit is in the region where the radial velocity data is concentrated, roughly $-5^\circ < \phi_1 < 15^\circ$. As a check, we perform an additional orbit fit using only the 6-D stars, which lie in the leading half of the stream near Segue 2 (see Figure \ref{fig:star_perturb}). This additional fit is very similar to the Fit Orbit, which demonstrates that the inferred orbit is not strongly affected by the choice to fit the entire stream vs. only stars closer to Segue 2's impact location.

The Tracer Orbit tracks the stream less well when compared to the Fit Orbit. As the CPS is a dwarf galaxy stream, it is relatively hot and we may expect that a single star may not accurately trace the entire stream, as seen here. However, the Tracer Orbit still broadly agrees with the shape of the stream, and fits quite well in the $-5^\circ < \phi_1 < 15^\circ$ region where radial velocity information is available. 

As another check on our stream member selection and orbits, we compare our results to the \textit{N}-body simulation of the CPS presented in \citet{chang_is_2020}, the present-day snapshot of which is made publicly available in \citet{yuan_complexity_2022}. While this simulation does not include the LMC, \citet{chang_is_2020} argue the LMC has a negligible effect on the past orbit of the CPS progenitor, as the LMC is far ($\sim 60$ kpc) from the CPS progenitor's last apocenter in their simulations, which is the only point where their orbits have comparable Galactocentric distances (see Sections \ref{ssubsec:MWLMC_mass} \& \ref{ssubsec:rigid_pots} for a further discussion of the LMC's effect on our results). 

Figure \ref{fig:sky_orbit} shows their present-day snapshot along with our member catalog and orbits on-sky. Our Fit Orbit agrees very well with the \citet{chang_is_2020} model, matching the position of the simulated stream over a full wrap. Our Tracer Orbit matches the simulation well in the South, but differs from the simulation by $\sim10^\circ$ in the North. The Fit (Tracer) orbit has an apocenter of $\sim 38$ (39) kpc and a pericenter of $\sim 20$ (16) kpc, which is also in broad agreement with the \citet{chang_is_2020} model and the orbits of individual CPS BHB stars in \citetalias{thomas_cetus-palca_2022}. 

Altogether, we are confident that our Fit Orbit is an appropriate model of the stream over the most recent wrap, while the Tracer Orbit provides an example of how individual stars move with respect to the overall stream track. Differences between these two orbits illustrate the uncertainties in modeling the stream, and will allow us to compare two different Segue 2 - CPS impact geometries moving forward in Section \ref{sec:segue2}.

\section{Interaction with Segue 2} \label{sec:segue2}

With the orbits of the CPS determined, we now seek to characterize the stream's interaction with Segue 2. To study whether Segue 2 had a close passage with the CPS, we use \texttt{gala} to compute the orbit of Segue 2 from its present-day position and velocity in our combined MW+LMC potential (see Section \ref{subsec:gals}). 

\begin{table}[]
    \caption{
    Summary of Segue 2's present-day position and velocity.
    } \label{tab:seg2_ic}
    \centering
    \begin{tabular}{c c c c}
        \hline
        \hline
         Parameter & Value & Unit & Reference \\
         \hline
         $\alpha$ & 34.8167 & deg & 1 \\
         $\delta$ & 20.1753 & deg & 1 \\
         Dist. modulus & 17.7$\pm$0.1 & mag & 1,2 \\
         $\mu_\alpha$ & 1.47$\pm$0.04 & mas yr$^{-1}$ & 3 \\
         $\mu_\delta$ & -0.31$\pm$0.04 & mas yr$^{-1}$ & 3 \\
         $v_r$ & -39.2$\pm2.5$ & km s$^{-1}$ &  1,2 \\
         \hline
         \hline
    \end{tabular}
    \tablenotetext{}{NOTE - Present-day phase space coordinates of Segue 2, including sky position, distance modulus, proper motions, and radial velocity. References: (1) \citealt{mcconnachie_revised_2020}; (2) \citealt{belokurov_discovery_2009}; (3) \citealt{mcconnachie_gaiaedr3pms_2020}.}
\end{table}

Segue 2's present-day 6-D phase space information is calculated from measurements of its radial velocity, distance modulus, and proper motions. We adopt the R.A., Decl., distance modulus, and heliocentric radial velocity from \citet{mcconnachie_revised_2020}, and the proper motions from \citet{mcconnachie_gaiaedr3pms_2020}, who compute proper motions of MW satellites using \textit{Gaia} EDR3. Note that the distance and radial velocity of Segue 2 were first reported in \citet{belokurov_discovery_2009}. To account for uncertainties in the present-day 3-D position and velocity vectors of Segue 2, we consider 2000 Monte-Carlo (MC) draws from the joint 1$\sigma$ error space of Segue 2's distance, proper motion, and radial velocity. These values are summarized in Table \ref{tab:seg2_ic}. From each MC draw, we derive Segue 2's 6-D phase-space vector in Galactocentric Cartesian coordinates before integrating its orbit. Our fiducial Segue 2 orbit is its ``direct orbital history,'' computed by taking the Table \ref{tab:seg2_ic} values without errors. 

\subsection{The Close Approach between Segue 2 and the Cetus-Palca Stream} \label{subsec:flyby}

\begin{figure*}
    \centering
    \includegraphics[width=\textwidth]{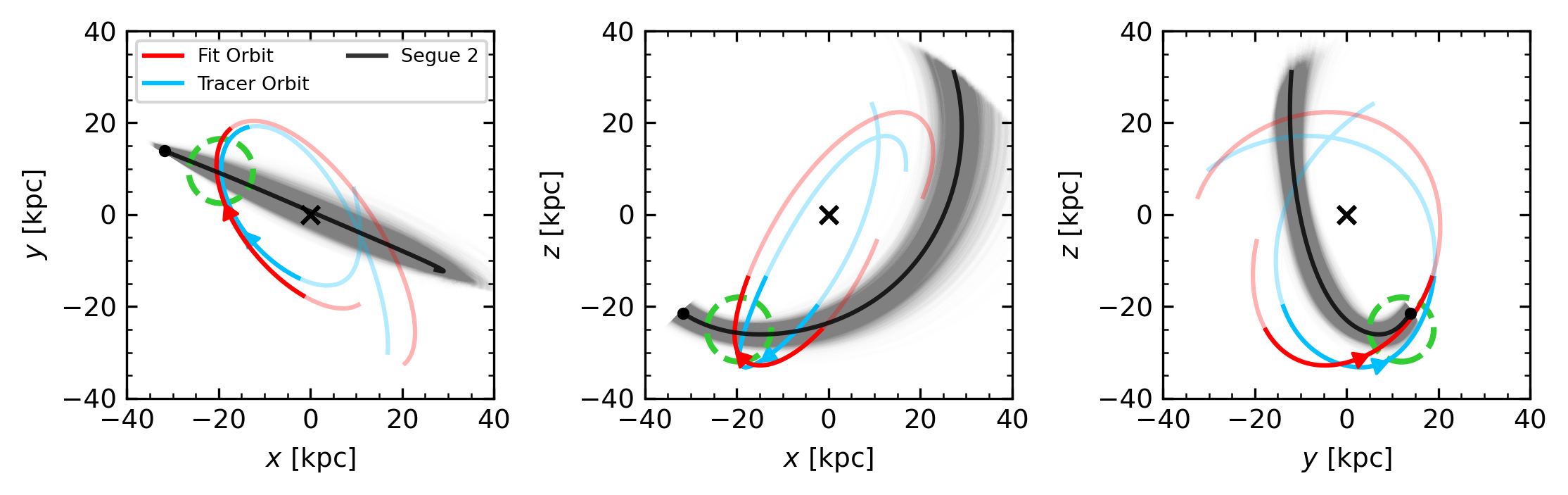}
    \caption{Galactocentric projections of our model stream orbits and Segue 2's orbit. Each panel shows a different projection, with the Galactic center marked by the cross. Our Fit CPS Orbit and Tracer CPS Orbit are shown as the red and blue lines, respectively. The bold sections of the stream orbits mark the portion that is covered by our member star catalog, and arrows denote the stream's direction of motion along the orbit. The present-day location of Segue 2 is marked by the black dot, and the black line shows its fiducial orbit over the past 500 Myr. The remainder of the MC Segue 2 orbits (as described in the text) are shown with gray lines. Segue 2 has a recent close approach with the CPS, the approximate location of which is marked with a dashed circle in each panel.}
    \label{fig:mc_orbits}

    \centering
    \includegraphics[width=\textwidth]{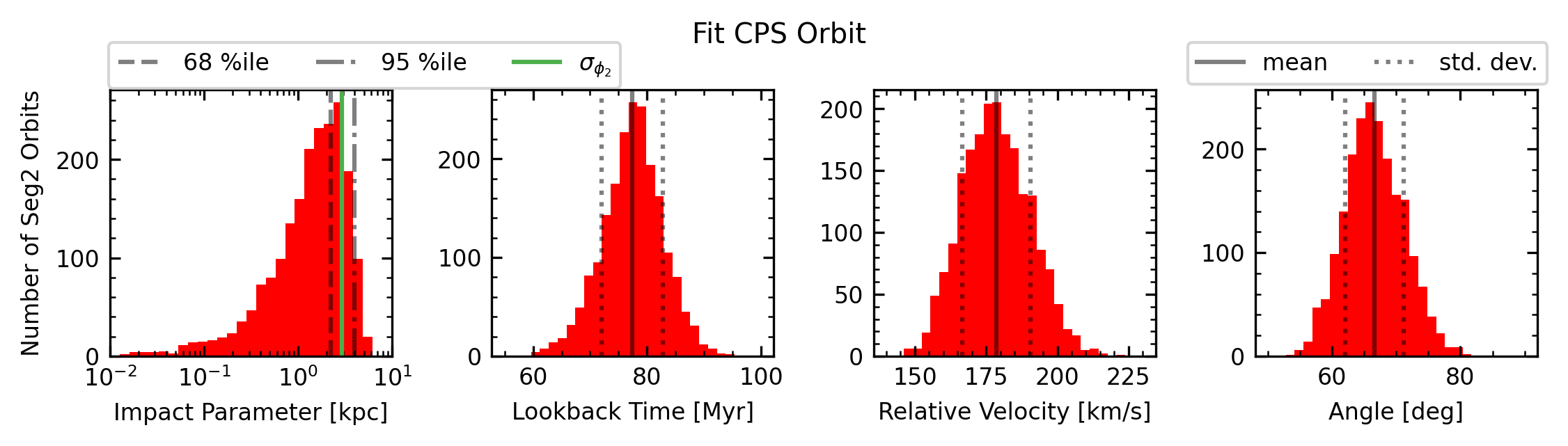}
    \includegraphics[width=\textwidth]{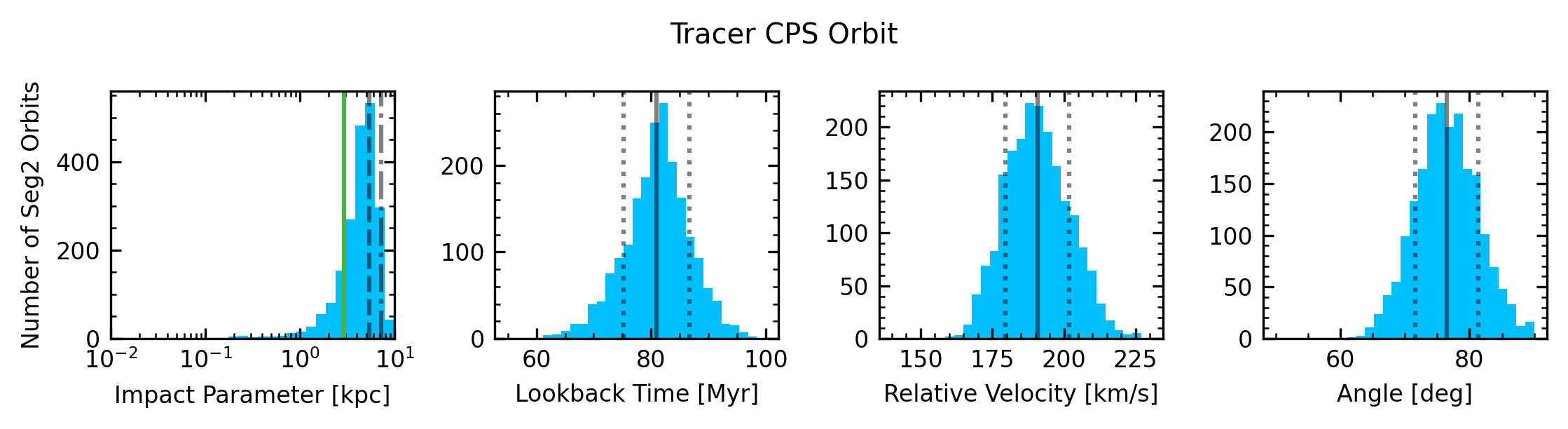}
    \caption{Distributions of the timings and geometry of the recent flyby encounter between the MC Segue 2 and our model orbits.
    The top (bottom) row of panels shows the orbit parameters at closest approach
    to the Fit CPS Orbit (Tracer CPS Orbit). From left-to-right, each row contains histograms of: the impact parameter or closest approach distance, lookback time or how far in the past the closest approach occurred, the relative velocity between Segue 2 and the stream test particle at the point of closest approach, and the angle between the Segue 2 orbits and the CPS orbit. A solid green line in the impact parameter panels denotes the 1$\sigma$ radius of the stream, while dashed and dot-dashed lines indicate the 68th and 95th percentile, respectively, of the Segue 2 orbits. The solid and dotted vertical lines in the other panels show the mean and standard deviation of the distributions. See Table \ref{tab:flybys} for a summary of these values. \textit{Every} Segue 2 orbit passes within 10 kpc of both model CPS orbits within 56-100 Myr ago. 
    }
    \label{fig:flyby_stats}

\end{figure*}

\begin{table}[]
    \caption{
    Summary of parameters describing the flyby between Segue 2 and the CPS.
    } \label{tab:flybys}
    \centering
    \begin{tabular}{c c c c}
        \hline
        \hline
         Parameter & Measured w/  & Measured w/  & Unit \\
          & Fit Orbit & Tracer Orbit & \\
         \hline
         \multicolumn{4}{c}{MC Segue 2 Orbits} \\
         \hline
         Med. Impact P. & 1.54 & 4.59 & kpc \\
         \% Seg2 $<$ 1$\sigma_{\phi_{2}}$ & 82.00 & 17.34 & - \\
         \% Seg2 $<$ 2$\sigma_{\phi_{2}}$ & 99.90 & 77.16 & - \\
         Lookback Time & $77.4 \pm 5.4$ & $81.0 \pm 5.7$ & Myr\\
         Rel. Velocity & $178.6 \pm 11.9$ & $190.8 \pm 11.1$ & km s$^{-1}$\\
         Angle & $66.6 \pm 4.6$ & $76.5 \pm 4.8$ & deg\\
         \hline
         \multicolumn{4}{c}{Fiducial Segue 2 Orbit} \\
         \hline
         Impact P. & 1.36 & 4.67 & kpc \\
         Lookback Time & 77.8 & 81.2 & Myr \\
         Rel. Velocity & 178.3 & 190.4 & km s$^{-1}$ \\
         Angle & 66.6 & 76.6 & deg \\
         \hline
         \hline
    \end{tabular}
    \tablenotetext{}{NOTE -In the top section of the table, impact parameter is reported as the median and percentage of MC Segue 2 orbits that pass within 1- and 2$\sigma_{\phi_2}$ (2.92 and 5.84 kpc) of the stream orbit as described in the text. Other reported values and uncertainties are the mean and standard deviation of the quantity derived from the 2000 MC Segue 2 orbits. The bottom section of the table lists the flyby parameters of the fiducial Segue 2 orbit.}
\end{table}

Figure \ref{fig:mc_orbits} shows the distribution of MC Segue 2 orbits (gray lines) with the fiducial present-day location and orbit of Segue 2 (black dot and line) highlighted. Each Segue 2 orbit has been integrated backwards in time for 500 Myr. The proper motion uncertainties are responsible for a $\sim 10^\circ$ spread in Segue 2's orbital plane, while the distance uncertainties drive a $\sim 10$ kpc spread in its pericenter distance. We also plot the model CPS orbits presented in Section \ref{subsec:orbit_results}, with the Fit and Tracer Orbit shown as the red and blue lines respectively. For each CPS orbit, we highlight the section of the orbit between $-56^\circ < \phi_1 < 38^\circ$ where our catalog of member stars is located. Arrows at arbitrary locations denote the direction the stream stars move along the orbits. 

All of the possible Segue 2 orbits pass close to the CPS orbits within the longitude range covered by stream stars (the flyby location is highlighted by a dashed green circle in each panel), implying that Segue 2 had a recent close flyby with the observable portion of the CPS. To facilitate future comparisons between this work and frameworks that recover encounter and perturber properties from stream perturbations \citep[e.g.,][]{erkal_properties_2015}, we quantify the probability, timing, and geometry of the flyby. Specifically, we calculate the minimum distance (impact parameter) between each MC Segue 2 orbit and both model CPS orbits. 
At each point of closest approach, we determine how far in the past the closest approach occurred (lookback time), the relative velocity between Segue 2 and the model CPS orbit's test particle at the time of closest approach, and the angle that Segue 2's trajectory makes with the stream orbit.

The distributions of these quantities are plotted in Figure \ref{fig:flyby_stats}. Each panel shows a histogram of one quantity for all of the 2000 MC Segue 2 orbits, measured with respect to one of the CPS model orbits. The top row is calculated with the Fit CPS Orbit, while the bottom row is calculated with the Tracer CPS Orbit. For lookback time, relative velocity, and angle, we include solid and dotted lines denoting the mean and standard deviation of the distribution. Numerical values of these statistics are provided in Table \ref{tab:flybys} along with the corresponding values for the fiducial Segue 2 orbit. 

For the impact parameters, the distributions are folded at zero (as negative impact parameters are impossible), so simply calculating and reporting the mean and standard deviation are not appropriate. Instead, we estimate the probability that Segue 2 has an impact parameter inside the stream as follows. We use $\sigma_{\phi_2} = 2.92$ kpc (the angular width of the stream in $\phi_2$ as described in Section \ref{subsec:fits}, converted to a length assuming a distance of 35 kpc) as an estimate for the width of the stream. The solid green line in the left column of panels denotes $\sigma_{\phi_2}$, while the dashed and dot-dashed lines in the impact parameter panels show the 68th and 95th percentile of the impact parameters. In our analysis, we consider impact parameters within $2\sigma_{\phi_2}=5.84$ kpc to be ``direct hits,'' i.e. Segue 2's center of mass passes within the radius from the stream's orbit that contains approximately 95\% of the stream stars. We refer to interactions with impact parameters larger than $\sigma_{\phi_2}$ as ``close passages.'' Table \ref{tab:flybys} reports the percentage of Segue 2 orbits with impact parameters less than $\sigma_{\phi_2}$ and $2\sigma_{\phi_2}$.

The median impact parameter is 1.54 kpc for the Fit Orbit, and 4.59 kpc for the Tracer Orbit. All but two of the 2000 possible Segue 2 orbits pass within $2\sigma_{\phi_2}$ of the Fit CPS Orbit, while only $77.16\%$ pass within $2\sigma_{\phi_2}$ of the Tracer CPS Orbit. This discrepancy is caused by the difference in the position of the model CPS orbits around $\phi_1 \sim 10^\circ$ where the interaction occurs (see Section \ref{subsec:perturbation}). In this region, the Tracer Orbit is at a larger distance and closer to the upper edge of the stream in $\phi_2$, slightly farther away from the fiducial Segue 2 orbit, which passes just inside the Fit Orbit (see Figure \ref{fig:mc_orbits}). 

The timing of the encounter is consistent between the stream orbit models and is roughly 80 Myr ago, while the relative velocity of the encounter differs at just over 1$\sigma$, with the Fit Orbit implying a $\sim10$ km s$^{-1}$ lower velocity ($\sim$180 km s$^{-1}$) than the Tracer Orbit ($\sim$190 km s$^{-1}$). 

Segue 2's orbit makes an angle of $\sim 67^\circ$ with the Fit CPS Orbit, and $\sim 76^\circ$ with the Tracer CPS Orbit. These differences in impact angle are barely more than 2$\sigma$, and are again driven by the Tracer Orbit being slightly less parallel to the stream stars in the vicinity of the flyby. 

In summary, we find that when considering the observational errors on Segue 2's distance and velocity vector, Segue 2 has a high probability (99.9\% or 77.16\% depending on the stream model) of directly impacting the CPS between 60-100 Myr ago.

\subsection{Perturbation Predictions} \label{subsec:perturbation}

\begin{table}[]
    \centering
    \caption{Summary of our Segue 2 models}
    \label{tab:segue2_models}
    \begin{tabular}{c c c c}
        \hline
        \hline
        Model & Profile & Total Mass & Scale Radius \\
         & & [$10^9 M_\odot$] & [kpc] \\
        \hline
        Seg2-1P & Plummer & 0.5 & 0.686 \\
        \textbf{Seg2-2P} & \textbf{Plummer} & \textbf{1.0} & \textbf{0.865} \\
        Seg2-3P & Plummer & 5.0 & 1.48 \\
        Seg2-3H & Hernquist & 5.0 & 8.35 \\
        Seg2-4P & Plummer & 10.0 & 1.86 \\
        \hline
        \hline
    \end{tabular}
    \tablenotetext{}{NOTE - For each model, we list the name, type of density profile, total mass, and scale radius. For a chosen mass and profile, we fit the scale radius such that each model is consistent with Segue 2's measured dynamical mass within its half-light radius from \citet{kirby_segue_2013} as described in the text. Our fiducial model (highlighted in bold) is Seg2-2P, with a mass of $10^9\,M_\odot$.}
\end{table}

In this section, we quantify the expected perturbation to the CPS due to Segue 2 as a function of Segue 2's mass and density profile. The Segue 2 models used in this section are summarized in Table \ref{tab:segue2_models}. As motivated in Section \ref{subsec:seg2_info}, these Segue 2 models are chosen to represent a range of possible formation scenarios for Segue 2, including tidally truncated halos of varying mass (represented by Plummer potentials) and a massive intact halo (represented with a Hernquist potential; note we reserve further discussion of Segue 2's density profile for Section \ref{ssubsec:density}). For each choice of profile and total mass, we fit the scale radius such that the mass enclosed within a radius of 46 pc is $1.5\times10^5 M_\odot$, consistent with the upper limit on Segue 2's dynamical mass within its half-light radius by \citet{kirby_segue_2013}. Throughout this section, we use Segue 2's fiducial orbit, i.e. the black line in Figure \ref{fig:mc_orbits}, unless otherwise noted. As a reminder, the lower portion of Table \ref{tab:flybys} lists the impact geometry for this orbit.

\begin{figure*}
    \centering
    \includegraphics[width=\textwidth]{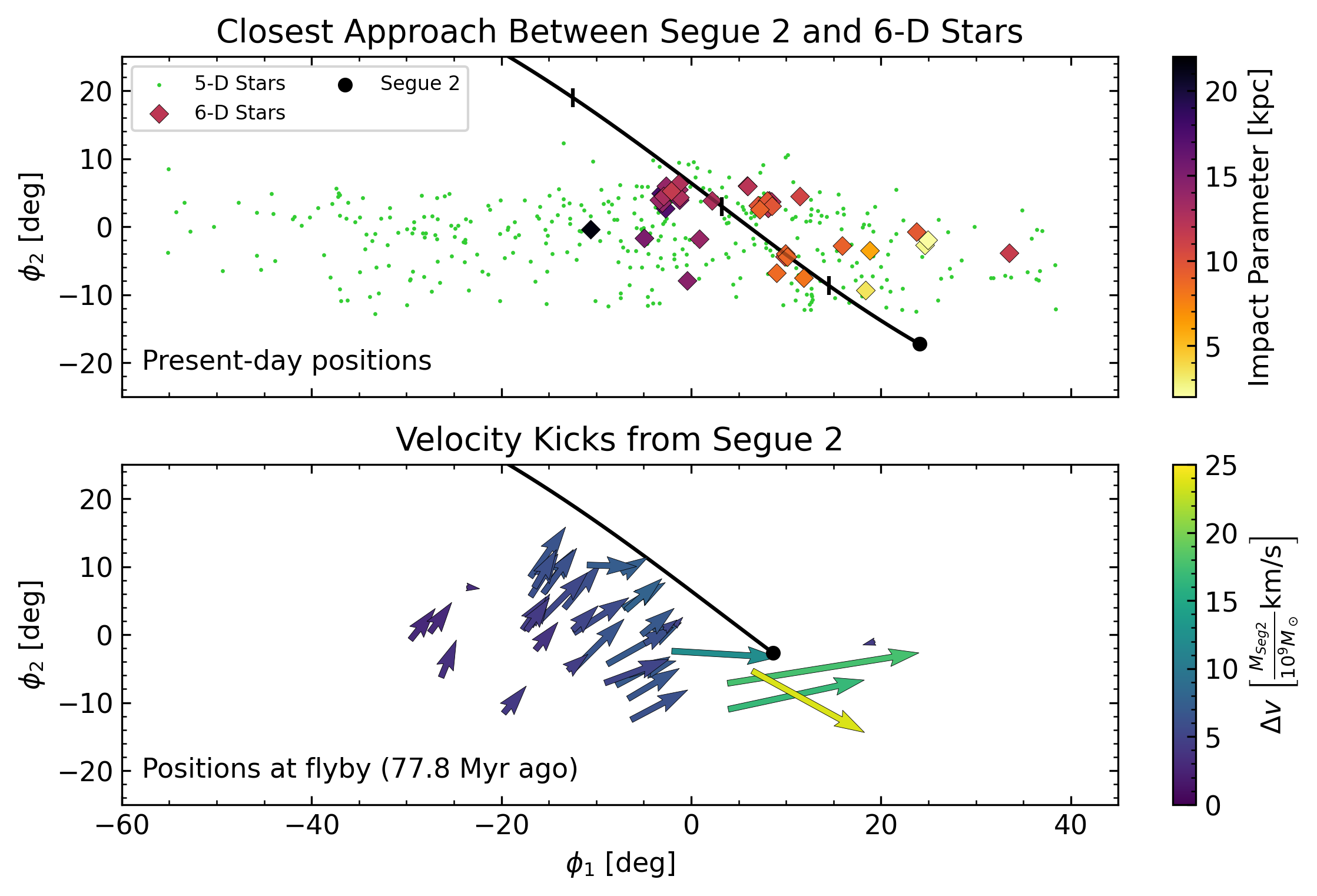}
    \caption{Geometry of the perturbation imparted on 6-D CPS stars from our Segue 2-2P model. The top panel shows the present-day locations of the 5-D stars (green points), 6-D stars (colored diamonds), Segue 2 (black point), and its fiducial orbit over the past 500 Myr (black line). Tick marks along Segue 2's orbit denote 50 Myr increments. 6-D star points are colored according to their minimum distance (impact parameter) to Segue 2 over the past 500 Myr. The bottom panel shows the same at the time of closest approach with the Fit CPS Orbit (77.8 Myr ago), omitting the 5-D stars as their positions at this time are not known. Arrows drawn from the 6-D stars' positions indicate the change in velocity of each star due to Segue 2 projected onto the plane of the sky, while the total (de-projected) magnitude of the kicks is denoted by the arrow colors. Segue 2 passes through the leading edge of our 6-D star sample, pulling stars near the flyby location primarily in the direction of their orbits, while deflecting the trailing portion of the stream slightly to the west.}
    \label{fig:star_perturb}
\end{figure*}

To begin, we integrate our 6-D CPS stars backwards in time for 500 Myr, adding the influence of our $10^9\,M_\odot$ Plummer Seg2-2P model such that the stars feel gravity from the MW, LMC, and Segue 2. For each star, we measure the closest approach / impact parameter between the star and Segue 2, as well as the change in the star's velocity due to Segue 2 (i.e. the difference in the star's velocity with and without Segue 2's gravity). The latter measurement is taken by numerically integrating over the star's acceleration from Segue 2 over the duration of the simulation. The results are plotted in Figure \ref{fig:star_perturb}. The top panel shows the locations of the 6-D stars and Segue 2 at the present day. We also include the 5-D stars for a reference of the position of the stream as a whole. The 6-D stars are colored by their impact parameter with respect to Segue 2. The stars around $\phi_1 \sim 25^\circ$ (similar to Segue 2's present longitude) have the smallest impact parameters of $\sim 2.5$ kpc, and are expected to be the most perturbed by Segue 2.

To show where the 6-D stars are relative to Segue 2 during the encounter, the bottom panel is drawn 77.8 Myr ago when Segue 2 has its closest approach with the Fit CPS Orbit. In this panel, we only include the 6-D stars (as the 5-D stars cannot be integrated backwards owing to their missing radial velocities). The 6-D stars are represented as arrows which point from the stars' positions and denote the magnitude and projected direction of their change in velocity due to Segue 2, based on integrating their accelerations from Segue 2 over the past 500 Myr. Assuming Segue 2 has a mass of $10^9\,M_\odot$, it changes the velocities of stars near the impact site ($\phi_1 \sim 10^\circ$ at 77.8 Myr ago) by $\sim 20$ km s$^{-1}$, comparable to the overall velocity dispersion of the stream. As the stars' accelerations due to Segue 2 are directly proportional to Segue 2's mass, the largest velocity perturbation received by a star from the Plummer models scales as $\Delta v_{\rm{max}} \sim (23.5 \, \rm{km\,s}^{-1})(M_{\rm{Seg2}}/10^9 M_\odot)$.\footnote{Note that in general, the perturbation strength also depends on the relative velocity, impact parameter, and scale radius of the perturber. Here, we assume the relative velocity is constrained (see Figure \ref{fig:flyby_stats}), the impact parameter is not significantly affected by Segue 2's mass, and that the change in Segue 2's scale radius with mass (see Table \ref{tab:segue2_models}) is small compared to the impact parameter. For a Hernquist model for Segue 2, the normalization changes to 4.7 km s$^{-1}$.} From this basic test, we can already see that it is likely that Segue 2 must be more massive than $10^9\,M_\odot$ to measurably increase the velocity dispersion of the CPS, a prediction which we refine with the models in Section \ref{subsec:models}.

There are insufficient numbers of stars with full 6-D data at present-day $\phi_1 > 20^\circ$ to reconstruct the interaction. As such, we construct a set of synthetic stream models to predict the signatures of Segue 2's flyby in the CPS as a function of Segue 2's mass and density profile. These models are described in Section \ref{subsec:models}, and the results are presented in Sections \ref{ssubsec:seg2_mass} (mass) and \ref{ssubsec:density} (density profile).

\subsection{Constraints on Segue 2 with Cetus-Palca Stream Models} \label{subsec:models}

The most detailed modeling strategy for the CPS would self-consistently generate a model stream by simulating the disruption a dwarf galaxy through particle-spray methods or an \textit{N}-body simulation (see e.g.,, \citealt{chang_is_2020}). Such an approach is expensive, however, especially when considering the need to fine-tune the initial conditions of the simulation to globally reproduce the observed properties of the stream. These difficulties are compounded by the fact that the CPS progenitor's location is unknown. In this section, our goal is to compare the perturbations generated by different Segue 2 mass models on the CPS. Therefore, we utilize a much less expensive strategy that involves distributing star particles along our stream orbits from Section \ref{subsec:orbit_results} according to the measured width of the stream in each dimension. 

In detail, we use both of our interpolated CPS orbits as ``templates'' for the synthetic streams that describe the streams' positions in $\phi_2$, $d$, $v_r$, $\mu_{\phi_1}$, and $\mu_{\phi_2}$ as a function of $\phi_1$ similarly to Section \ref{subsec:fits}. For each synthetic stream, we distribute 1000 star particles uniformly in $-80^\circ \leq \phi_1 \leq 70^\circ$, with the other five phase-space coordinates specified by the template orbit. This results in a linear density of stars similar to the most dense portion of the observed stream. 

We then introduce a scatter in the star particles' phase-space positions, drawing the scatter in each coordinate from an independent normal distribution. The width of the distribution in each dimension is as follows. The spread in $\phi_2$ is $\sigma_{\phi_2}$. The scatter in each proper motion direction is estimated with the same procedure as $\sigma_{\phi_2}$ (see Section \ref{subsec:fits}), i.e. we calculate the binned standard deviation of the stream in $\mu_{\phi_2}$ and then take the mean of the binned widths. In principle, the width of the observed stream already encodes perturbations from Segue 2, but as only a portion of the stream is near the impact site, averaging over longitude bins reduces the contribution of the perturbed portion to the width measurement. The spread in $d$ ($v_r$) is the same as the spread in $\phi_2$ $(\mu_{\phi_2}$), translated to a length (linear velocity) using the distance of the template orbit at the $\phi_1$ location of each star. This process generates a population of stars that locally reproduces the observed properties of the CPS, suitable for estimating Segue 2's influence on the stream.

\begin{figure*}
    \centering
    \includegraphics[width=1.0\textwidth]{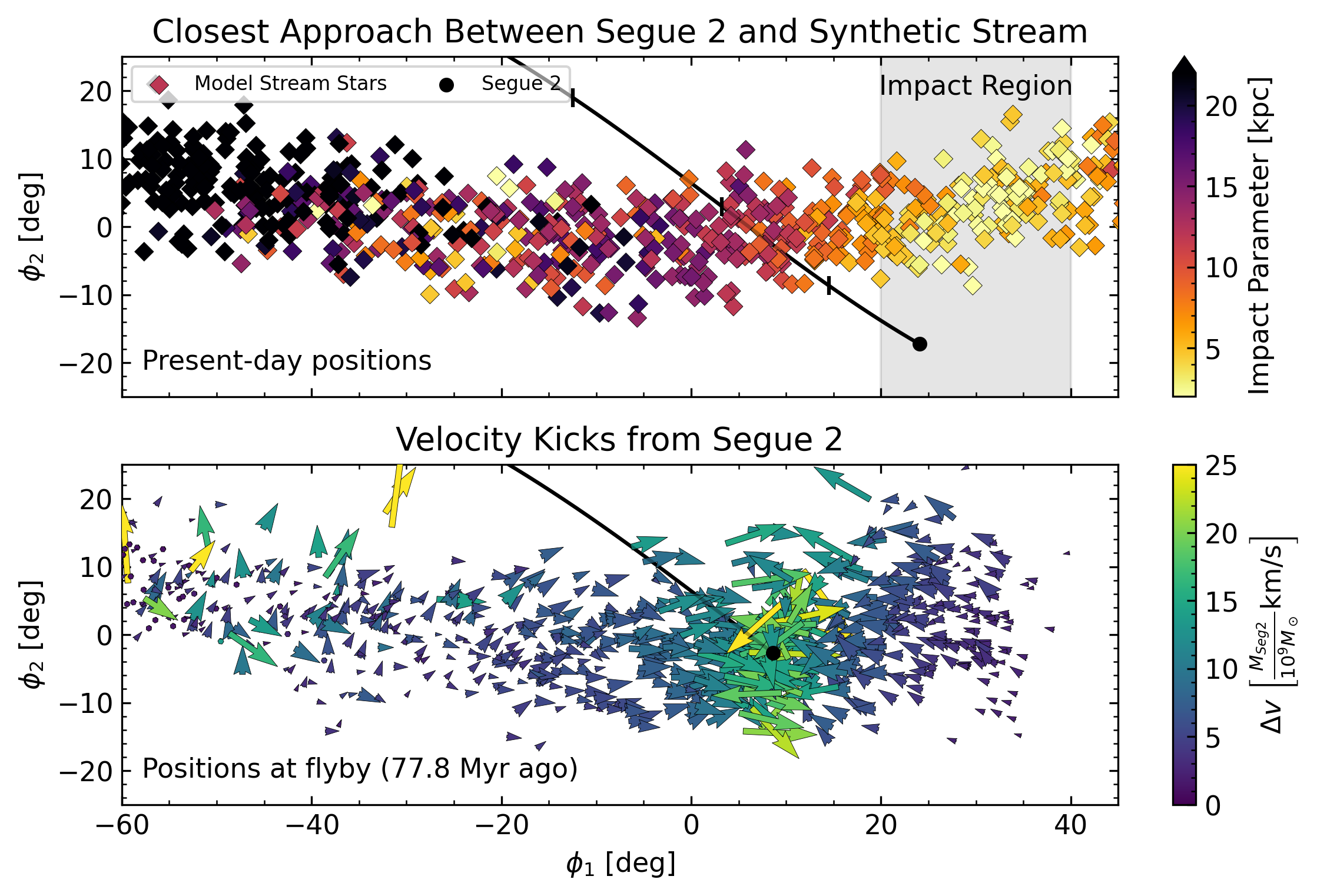}
    \caption{Same as Figure \ref{fig:star_perturb}, but showing the synthetic stream generated from the Fit Orbit template. Note that we keep the color scales identical to Figure \ref{fig:star_perturb} to ease comparison, though there are many stars in the trailing arm of the stream with impact parameters greater than 20 kpc. In the top panel, we see that the stars with the lowest impact parameters are between $20^\circ < \phi_1 < 40^\circ$ (the shaded Impact Region) in agreement with our sparse 6-D data in Figure \ref{fig:star_perturb}. A few stars between $-40^\circ \lesssim \phi_1 \lesssim -15^\circ$ also come within 5 kpc of Segue 2 approximately half an orbital period ago, where Segue 2 passes near (but not through) the CPS again (see Figure \ref{fig:mc_orbits}). The bottom panel also agrees with Figure \ref{fig:star_perturb}, as we see Segue 2 deflecting stars behind (ahead of) the impact location towards positive (negative) $\phi_1$. Ultimately, our models predict that most CPS stars at $20^\circ < \phi_1 < 40^\circ$ have passed within $\sim$5 kpc of Segue 2, which underscores the need for more radial velocity measurements of CPS stars with positive $\phi_1$ to confirm this.
    }
    \label{fig:model_perturb}
\end{figure*}

We integrate the star particles from the synthetic streams backwards in time for 500 Myr in the gravity of the MW and LMC only. We then inject Segue 2 on its fiducial orbit and integrate forward to the present day, allowing the stars to feel the gravity of all three galaxies. Figure \ref{fig:model_perturb} shows the synthetic stream generated from the Fit Orbit template and perturbed by the Segue 2-2P model in the same fashion as Figure \ref{fig:star_perturb}. The top panel illustrates that Segue 2 passed closest to the portion of the model stream that is just ahead of Segue 2 at the present day ($20^\circ < \phi_1 < 40^\circ$; which we refer to as the Impact Region). The lower panel illustrates that at closest approach, Segue 2 imparted velocity kicks of $\sim$20 km s$^{-1}$ to these stars. While these velocity kicks converge towards Segue 2, i.e. kicks at $\phi_1 > 8.5^\circ$ point to the left, and kicks at $\phi_1 < 8.5^\circ$ point to the right, the largest effect is to randomize the velocity field at the impact site itself.  

While this picture is consistent with our observational data in Figure 9, there are only four stars in our dataset in the Impact Region ($20^\circ < \phi_1 < 40^\circ$) with measured radial velocities (out of 39), limiting detailed comparisons to our models. Figure 10 demonstrates that constraints on Seque 2’s DM profile are possible if additional radial velocity data is obtained for stars in the Impact Region. 

Interestingly, in Figure \ref{fig:model_perturb} we also see a small number of stars between $-40^\circ \lesssim \phi_1 \lesssim -15^\circ$ with small impact parameters and large velocity kicks. This likely corresponds to a previous close approach between Segue 2 and the CPS orbit $\sim$half an orbital period ago ($\sim$400 Myr ago; see Figure \ref{fig:mc_orbits}). While this earlier encounter is weaker than the recent encounter we study in this paper, this raises the intriguing possibility that Segue 2 has been affecting the CPS for much longer than the past 100 Myr. Radial velocity data for CPS stars at $\phi_1 < -15^\circ$ are needed to compute their orbits and verify this idea.

\subsubsection{The Mass of Segue 2} \label{ssubsec:seg2_mass}

\begin{figure*}
    \centering
    \includegraphics[width=\textwidth]{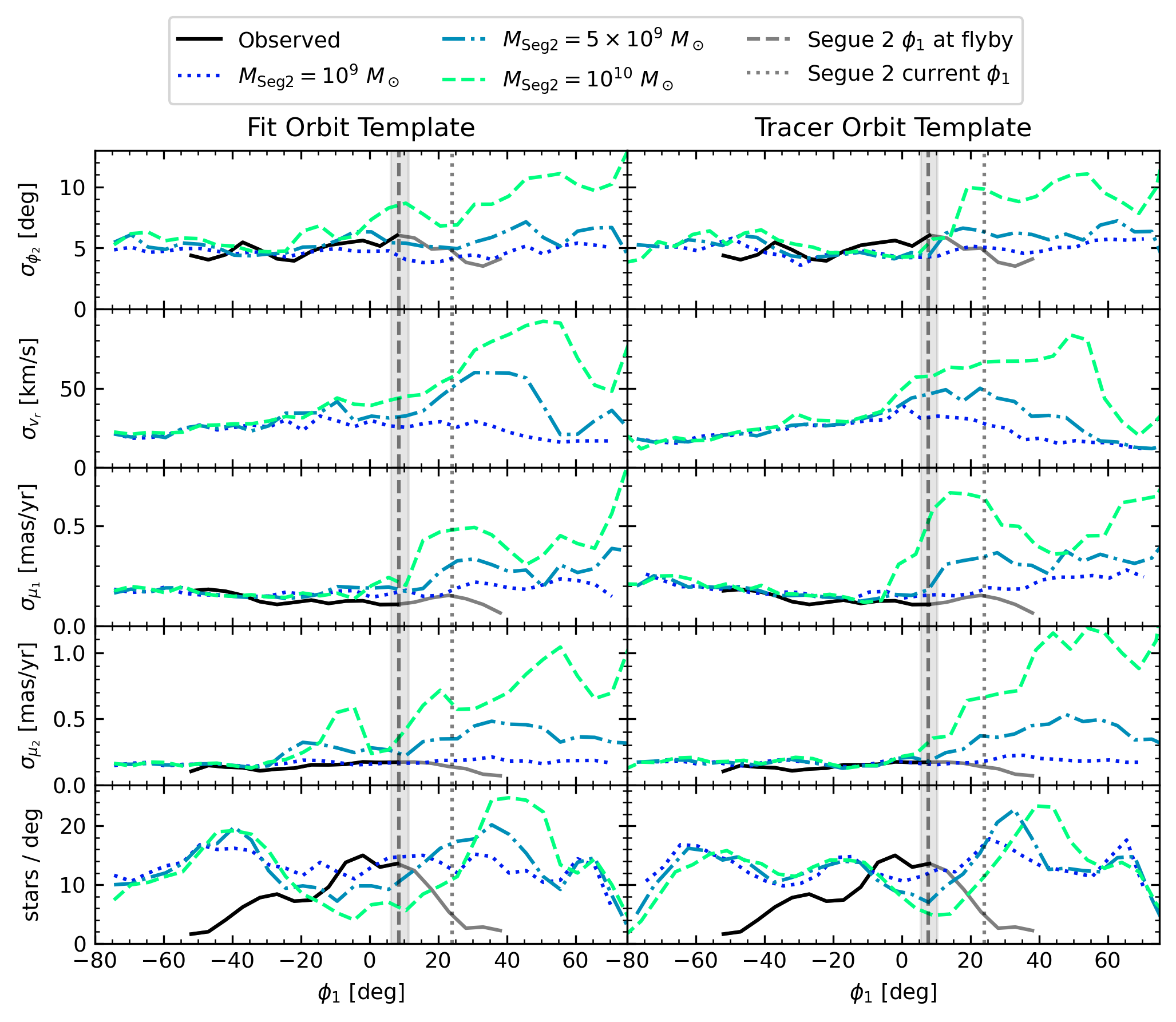}
    \caption{Width and density of our synthetic streams as a function of $\phi_1$ and Segue 2's mass. The left column of panels shows the streams generated from the Fit Orbit, while the right column shows streams generated from the Tracer Orbit. Each panel shows the width of the observed stream (solid black line), and perturbed models (dotted, dash-dotted, and dashed lines) in one dimension, with the bottom row showing the linear density of the streams. In all panels, we reduce the opacity of the observed stream's line for $\phi_1 \geq 10^\circ$ to indicate where direct comparisons to the models become unreliable, as explained in the text. The streams perturbed by a $10^9$ $M_\odot$ Segue 2 are effectively identical to the $5\times10^8$ $M_\odot$ and unperturbed cases, so we omit the latter two for clarity. Note that we do not include the observed radial velocity width of the CPS as we have too few stars with measured radial velocities to plot a width reliably along the full length of the stream. For reference, vertical dotted lines indicate the present-day $\phi_1$ location of Segue 2, and vertical dashed lines show the location of closest approach of the fiducial Segue 2 orbit to the stream orbit. A shaded region shows the spread of the flyby locations from the 2000 MC Segue 2 orbits described in Section \ref{subsec:flyby}. Significant heating (larger than the shot noise from placing the stars) of the CPS is expected at $\phi_1 \gtrsim 10^\circ$ if Segue 2 has a mass $> 10^9 M_\odot$. A Segue 2 mass of $10^{10} \, M_\odot$ is inconsistent with the observed properties of the CPS at $\phi_1 < 0^\circ$. }
    \label{fig:mockstreams}
\end{figure*}

In this section, we use the different mass models for Segue 2 (see Table \ref{tab:segue2_models}) to investigate how Segue 2's effect on the CPS depends on Segue 2's mass. 
In the previous section, we used a synthetic stream encounter with Segue 2 to demonstrate that the primary kinematic signature of Segue 2’s flyby will be heating in the stream in the Impact Region ($20^\circ < \phi_1 < 40^\circ$ at the present day; see lower panel of Figure \ref{fig:model_perturb}).
To quantify the heating of the stream by Segue 2, we measure the dispersion of the stream in $\phi_2$ and its three velocity components in a series of $10^\circ$ - wide $\phi_1$ bins, shifted by $5^\circ$ along the stream (i.e. the bin centers are separated by $5^\circ$ such that each $5^\circ$ segment of the stream is included in two adjacent bins). In addition, we measure the linear density of the stream by counting the number of stars in each bin. 

We repeat this process for each of the synthetic stream templates (CPS orbits) and each of the Plummer Segue 2 models in Table \ref{tab:segue2_models}. The results for the streams generated from the Fit and Tracer Orbits are shown in the left and right columns of Figure \ref{fig:mockstreams}, respectively. For comparison, we also include the width / density of the observed CPS in every dimension except radial velocity dispersion, as we do not have enough stars with spectroscopic radial velocities to accurately measure this quantity. To guide the eye, a dashed vertical line in each panel denotes the $\phi_1$ location of the fiducial Segue 2 orbit's closest approach to the stream orbit, while the shaded region shows the 2.5 - 97.5 percentile range of the MC Segue 2 flyby locations. Similarly, a dotted vertical line shows the present-day $\phi_1$ location of Segue 2. Note that we omit the $M_{\rm{Seg2}}=5\times10^8\,M_\odot$ case and the unperturbed synthetic stream for clarity, as these are effectively identical to the $M_{\rm{Seg2}}=10^9\,M_\odot$ case (dotted lines). 

In other words, if Segue 2's mass is $\leq 10^9\,M_\odot$, it does not leave a detectable perturbation signature in the dispersions of the stream stars. This agrees with the result in Section \ref{subsec:perturbation} that the velocity kicks imparted to stream stars by a $10^9\,M_\odot$ Segue 2 are comparable to the intrinsic dispersion of the stream. 

Conversely, if Segue 2 has a mass of $5 \times 10^9\,M_\odot$, it noticeably heats the stars in the vicinity of the flyby. By the present day, the proper motion (radial velocity) dispersion of the stream near Segue 2's present-day location ($\phi_1 \approx 25^\circ$, within the Impact Region) is $\sim 0.15$ mas yr$^{-1}$ (40 km s$^{-1}$) higher than the unperturbed portion of the stream. In the $M_{\rm{Seg2}}=10^{10}\,M_\odot$ case, the heating is even larger, $\sim 0.3$ mas yr$^{-1}$ in proper motions or $\sim$ 50 km s$^{-1}$ in radial velocity in the vicinity of Segue 2. 

Notably, a $10^{10}\,M_\odot$ Segue 2 also noticeably heats the stream \textit{behind} the flyby location. In particular, the $\mu_{\phi_2}$ dispersion of the stream based on the Fit Orbit is $\sim0.3$ mas yr$^{-1}$ higher around $\phi_1 \sim -10^\circ$ in response to a $10^{10}\,M_\odot$ Segue 2 than the other masses we consider. A similar bump in the $\mu_{\phi_1}$ dispersion near $\phi_1 \sim 0^\circ$ can be seen in the stream based on the Tracer Orbit in the presence of a $10^{10}\,M_\odot$ Segue 2. It is possible that Segue 2's previous approach to the CPS (see Section \ref{subsec:models}) contributes to the heating at $\phi_1 < 0$, though such an effect is inconsistent with the observed CPS.

Direct comparisons between the simulated and observed velocity dispersions at $\phi_1 \gtrsim 10^\circ$ are challenging as the density of the CPS drops quickly with increasing $\phi_1$ beyond this point (bottom panels of Figure \ref{fig:mockstreams}). The series of cuts we employed in the construction of the catalog (Section \ref{sec:obs}) may exclude stream stars that have experienced large velocity kicks from Segue 2, which may artificially drop the CPS density in this region.

Alternatively, the drop in density may be a genuine gap caused by Segue 2. To investigate this possibility, we compare the observed decrease in density to the simulated streams (bottom panels of Figure \ref{fig:mockstreams}). A $10^9\,M_\odot$ Segue 2 does not cause significant variations in the density of our synthetic streams above the shot noise from placing the star particles. For higher Segue 2 masses however, the beginning stages of gap formation can be seen in the form of an underdensity around $\phi_1 \sim 0^\circ$ and corresponding overdensity around $\phi_1 \sim 30^\circ$. This effect is mild in the case of a $5 \times 10^9\,M_\odot$ Segue 2. However, if Segue 2 has a mass of $10^{10}\,M_\odot$, the significant underdensity in the simulated streams at $\phi_1 \sim 0^\circ$ is inconsistent with the observed CPS's density \textit{peak} at this location. 

Running all three simulations into the future reveals that a gap in the stream due to Segue 2 forms $\sim 300$, 200, or 70 Myr from the present day if Segue 2's mass is $10^{9}$, $5\times10^9$, or $10^{10}\,M_\odot$, respectively. This implies that for a $10^{10}\,M_\odot$ Segue 2 to cause a gap by the present day, the impact would have to occur 70 Myr earlier than the fiducial encounter we study in this section, i.e. $(77.8+70)\,\rm{Myr}=147.8$ Myr ago. However, the encounter must occur no more than 100 Myr ago based on the allowed Segue 2 orbits (see Figure \ref{fig:flyby_stats}). Therefore, even the earliest allowable flyby of the most massive Segue 2 model we consider leaves insufficient time for gap formation before the present day. 

Instead, we suggest three potential causes of the observed drop in density at the leading edge of the stream: 
\begin{enumerate}
    \item Incompleteness and/or crowding in the observations in this region as the stream approaches the MW's disk.
    \item Intrinsic variations in the density of the stream due to the details of its formation process (e.g.,, epicyclic density variations; \citealt{kupper_structure_2008, just_quantitative_2009, kupper_tidal_2010, kupper_more_2012, mastrobuono-battisti_clumpy_2012}).
    \item Selection effects from our catalog construction process (see Section \ref{sec:obs}). If the stream is indeed heated by Segue 2 at $\phi_1 \gtrsim 10^\circ$, a flat proper motion cut such as the one employed here would select fewer stars in the heated portion of the stream.
\end{enumerate}

\noindent Future work will be needed to determine the contribution of each effect. 

To summarize, we find that a Segue 2 mass as high as $10^{10}\,M_\odot$ produces high-level inconsistencies between the observed and simulated properties of the stream. We cannot confidently rule out a mass of $5\times10^9 M_\odot$ given the outstanding uncertainties in the observations and the inherent limitations of our simplistic modeling approach. If Segue 2's mass is $\leq 10^9\,M_\odot$, we should not expect future observations to see heating in the CPS due to the interaction at present. None of the Segue 2 models or orbits we consider are capable of creating a density gap in the CPS by the present day. The primary effect of a sufficiently massive Segue 2 at present is heating of the stream in all three velocity components at $\phi_1 \gtrsim 10^\circ$, with the largest effects in the Impact Region ($20^\circ < \phi_1 < 40^\circ$).

\subsubsection{The Density Profile of Segue 2}\label{ssubsec:density}

\begin{figure}
    \centering
    \includegraphics[width=\columnwidth]{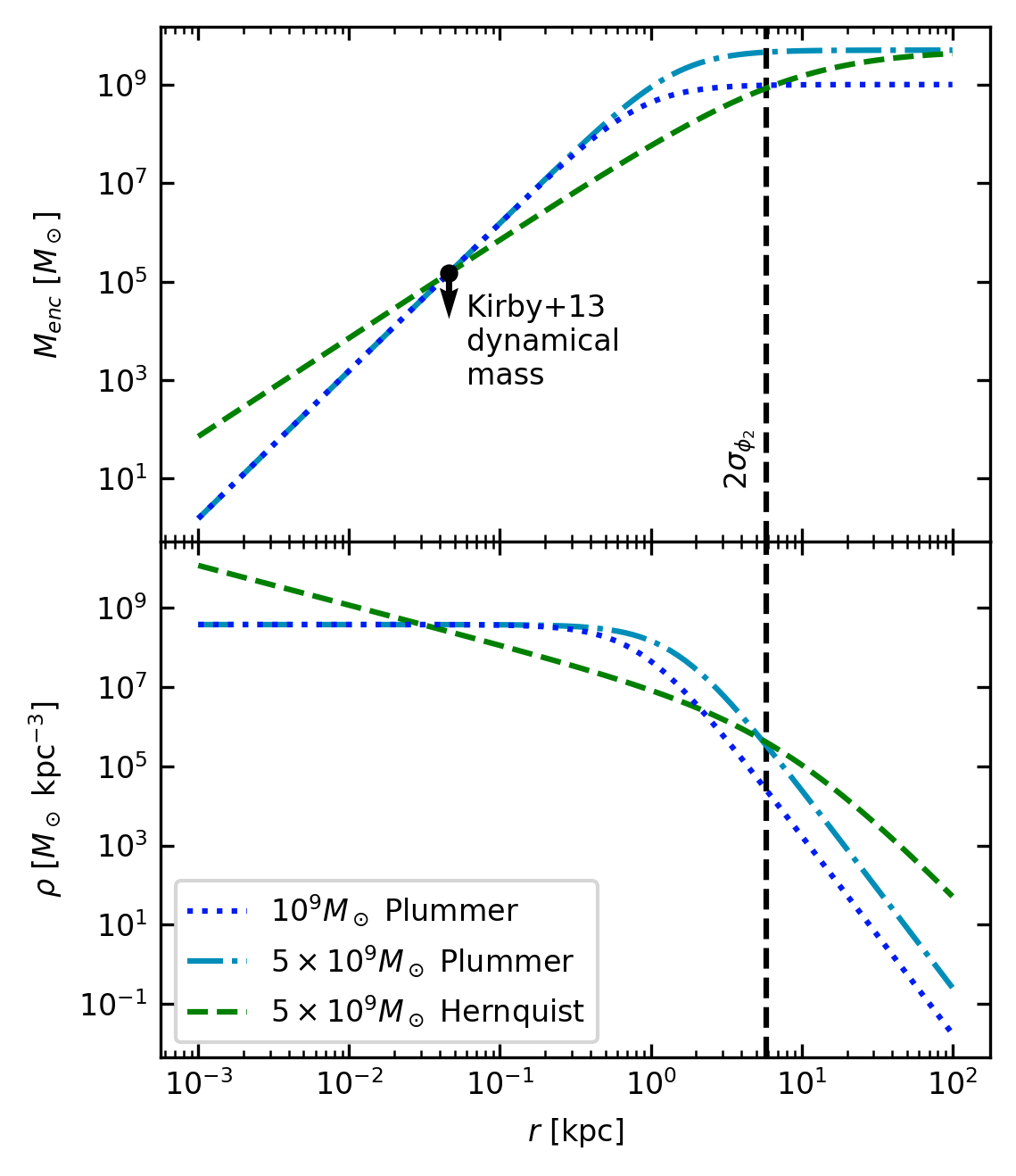}
    \caption{Mass (top) and density (bottom) profiles for the Segue 2 models studied in Section \ref{ssubsec:density}. The $10^9\,M_\odot$ and $5\times10^9\,M_\odot$ Plummer profiles are shown as the dotted and dash-dotted lines, while the $5\times10^9\,M_\odot$ Hernquist profile is shown as the dashed line. All profiles match the \citet{kirby_segue_2013} dynamical mass limit for Segue 2 by construction. A vertical dashed line indicates the 2$\sigma$ width of the CPS, which marks the radius from Segue 2's center that the stream is sensitive to Segue 2's mass as explained in the text.}
    \label{fig:m_profs}
\end{figure}

Here, we investigate the possibility of using the Segue 2 - CPS interaction to constrain the density profile of Segue 2's DM halo. As a basic test, we compare our Plummer sphere models to a Hernquist sphere with a total mass of $5\times10^9\,M_\odot$
(see Section \ref{ssubsec:seg2_mass} \& Table \ref{tab:segue2_models}). We choose this mass as it is above the threshold where the CPS is sensitive to Segue 2 in Figure \ref{fig:mockstreams}, allowing us to compare the effects of a cuspy (Hernquist) vs. cored (Plummer) profile. Note that in addition to having a central core, a Plummer profile falls off more steeply than a Hernquist profile outside of its scale radius, and is therefore a better approximation of a tidally truncated halo. For reference, we plot the mass and density profiles of the $10^9\,M_\odot$ and $5\times10^9\,M_\odot$ Plummer profiles and the $5\times10^9\,M_\odot$ Hernquist profile in Figure \ref{fig:m_profs}. As a reminder, the enclosed mass of all profiles is $1.5\times10^5\,M_\odot$ at a radius of 46 pc (\citealt{kirby_segue_2013}; black circle) by construction. 

The Hernquist model for Segue 2 leaves no detectable perturbation in the width or density of either synthetic stream by the present day. In other words, if we were to plot the properties of the synthetic stream perturbed by the Hernquist Segue 2 model on Figure \ref{fig:mockstreams}, it would look extremely similar to the synthetic stream perturbed by the $M_{\rm{Seg2}}=10^9\,M_\odot$ Plummer sphere Segue 2 model. 

This can be understood in terms of the mass enclosed by each Segue 2 model within its impact parameter to the stream stars,
which determines the gravitational acceleration felt by the stars for a fixed Segue 2 orbit.  
Depending on Segue 2's impact parameter to the stream's orbit, the encounters allowed by the MC Segue 2 orbits (see Figure \ref{fig:flyby_stats}) can be split into two cases according to the definitions in Section \ref{subsec:flyby}:

\begin{itemize}
    \item Direct hits (impact parameters in (0, $2\sigma_{\phi_2}\approx6$] kpc): In this case, Segue 2's center of mass passes within the stream. \textit{Individual} stream stars will therefore have impact parameters with respect to Segue 2 spanning the width of the stream, i.e. within (0, $2\sigma_{\phi_2}$]. 
    \item Close passages (impact parameters between 6 - 10 kpc; note that no Segue 2 orbits allowed within the error space have impact parameters larger than 10 kpc): In this case, some \textit{individual} stream stars will still pass within $2\sigma_{\phi_2}$ of Segue 2's center of mass, because Segue 2 cannot be more than $10 - 5.84 = 4.16$ kpc from the edge of the stream. 
\end{itemize}

Therefore, within the range of impact parameters to the CPS orbit allowed by Segue 2, individual stream stars will probe Segue 2's mass within roughly $2\sigma_{\phi_2}\approx6$ kpc. In Figure \ref{fig:m_profs}, we mark $2\sigma_{\phi_2}$ with a dashed, vertical line. Note that within this distance, the enclosed mass of the $10^9\,M_\odot$ Plummer profile is very similar to the $5\times10^9\,M_\odot$ Hernquist profile, explaining why both models have the same effect on the width of our synthetic streams. 

This allows us to refine our conclusions from the previous section about Segue 2's mass profile. Namely, \textit{any} Segue 2 DM density profile that encloses a total mass $\leq 10^9 M_\odot$ within $\sim 6$ kpc is not expected to leave a measurable perturbation on the CPS. On the other hand, if a perturbation is detected, this would put a lower limit on Segue 2's mass within $\sim 6$ kpc of $> 10^9 M_\odot$. 

Crucially, the Segue 2 - CPS interaction is unique in that the perturbing DM halo hosts a luminous galaxy that has a dynamical mass estimate within $\sim$ 50 pc. When combined with a mass measurement at kiloparsec scales from our predicted stream perturbation, these two data points will measure the slope of Segue 2's density profile. 

For the sake of a strong example, let us pretend that a future observation with a well-characterized selection function has searched for and discovered a hot component of the CPS in the Impact Region which is consistent with our predictions for a $5\times10^9\,M_\odot$ Plummer sphere (i.e. Segue 2's mass is $5\times10^9\,M_\odot$ within $\sim6$ kpc). In this hypothetical case, to simultaneously satisfy both this constraint and the \citet{kirby_segue_2013} dynamical mass constraint at 46 pc, Segue 2's density profile could not fall off more steeply than approximately $r^{-0.86}$ over this range, which would exclude an NFW or Hernquist profile, though not a DM cusp. In short, Segue 2's interaction with the CPS may be capable of providing a second data point on the mass profile of a UFD at a drastically different distance scale than is probed by its internal dynamics. 

\begin{figure}
    \centering
    \includegraphics[width=\columnwidth]{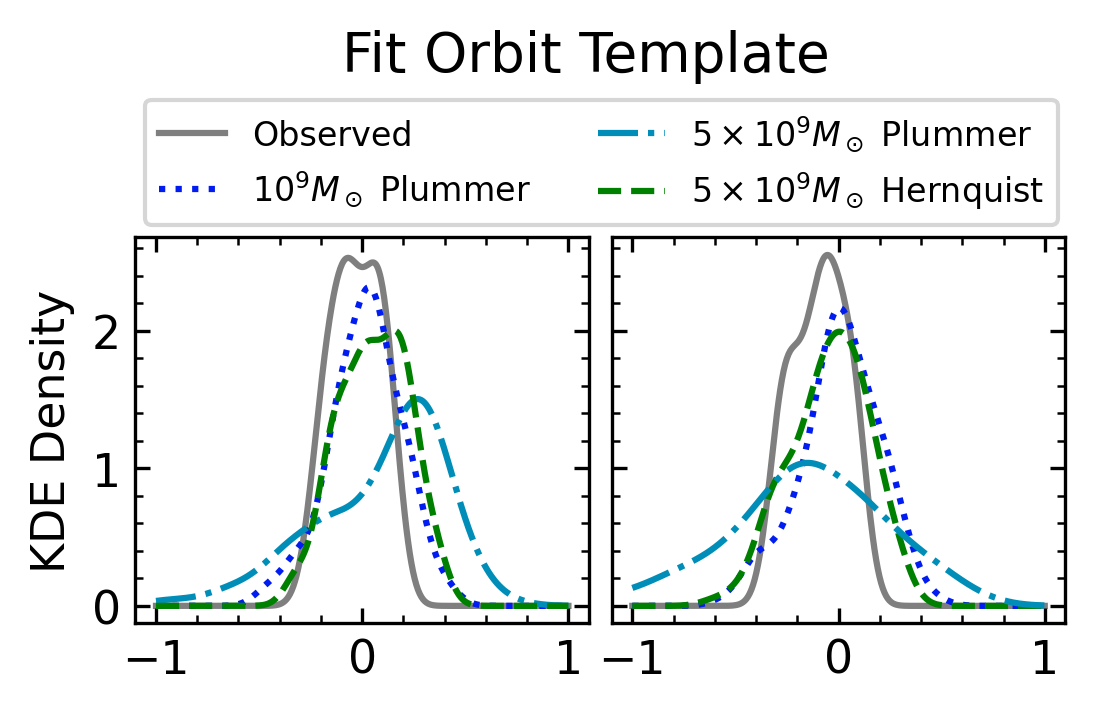}
    \includegraphics[width=\columnwidth]{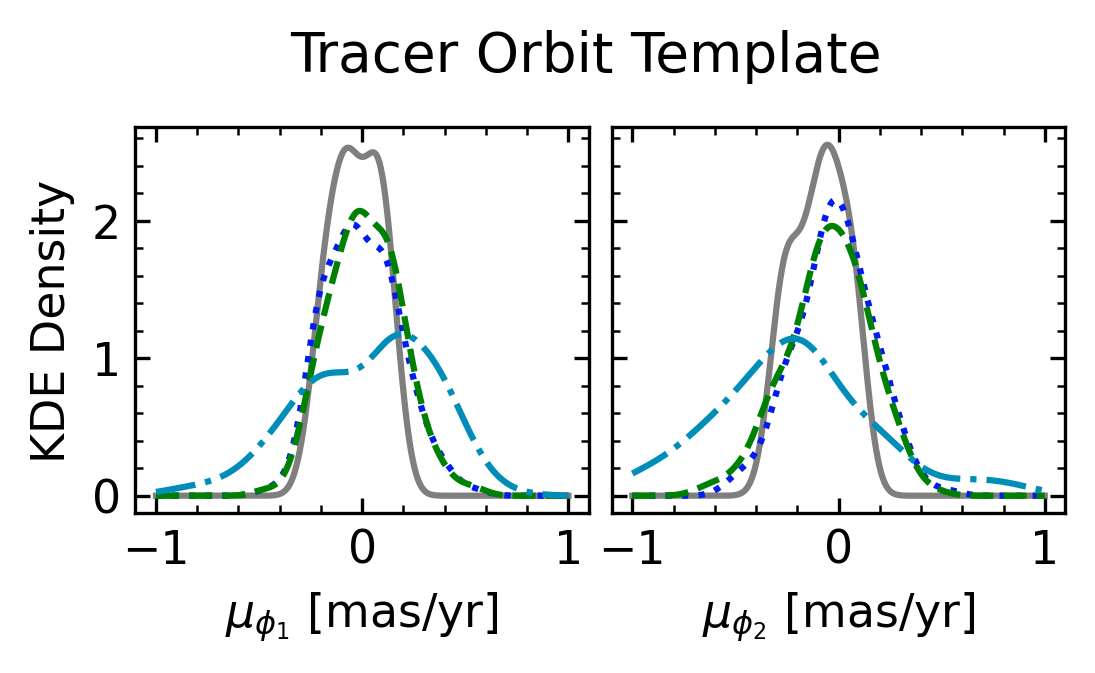}
    \caption{Proof of concept for using the velocity distribution of stream stars to distinguish between Segue 2 models. Each panel shows the proper motion distributions of synthetic stream stars within $20^\circ < \phi_1 < 40^\circ$ (the impact site; see Figure \ref{fig:model_perturb}) as broken, colored lines. The top (bottom) row shows the synthetic streams generated from the Fit (Tracer) Orbit, as well as the observed CPS stars. For both synthetic stream templates, the $5\times10^9\,M_\odot$ Plummer Segue 2 broadens the tails of the distributions in both components, and shifts the peak of the distribution by roughly $-0.2$ mas yr$^{-1}$ in $\mu_{\phi_1}$ and $0.2$ mas yr$^{-1}$ in $\mu_{\phi_2}$ when compared to the $10^9\,M_\odot$ Plummer case. For the Tracer Orbit synthetic streams, the $5\times10^9\,M_\odot$ Hernquist case is similar to the $10^9\,M_\odot$ Plummer case. However, for the Fit Orbit synthetic streams, the $5\times10^9\,M_\odot$ Hernquist Segue 2 creates a shoulder in the $\mu_{\phi_2}$ distribution near $-0.3$ mas yr$^{-1}$ as well as a second peak in the $\mu_{\phi_1}$ distribution near $0.2$ mas yr$^{-1}$, which are not created by the $10^9\,M_\odot$ Plummer profile. As a proof of concept, the proper motion distributions for the current data (gray lines) are also included. While direct comparisons between the data and synthetic streams are not yet possible (as explained in the text), the observed $\mu_{\phi_2}$ distribution has a shoulder that is similar to those seen in the models, demonstrating that the velocity distribution of stars near the impact site can in principle distinguish between Segue 2 mass profiles.}
    \label{fig:vDists}
\end{figure}

We next examine the impact of changes in Segue 2's DM density profile on the velocity distribution of CPS stars. We plot KDEs of the proper motion distributions of stars in the Impact Region ($20^\circ < \phi_1 < 40^\circ$) for the synthetic streams perturbed by our Seg2-2P, Seg2-3P, and Seg2-3H Segue 2 models (those shown in Figure \ref{fig:m_profs}) in Figure \ref{fig:vDists}. In this figure, the proper motions of the synthetic stream stars are shown with respect to the orbit template used to generate the synthetic stream. The $5\times10^9\,M_\odot$ Plummer model for Segue 2 has the largest effect on the velocity distributions, broadening the tails and shifting the peak of the distribution up (down) by about 0.2 mas yr$^{-1}$ in $\mu_{\phi_1}$ ($\mu_{\phi_2}$). This shifting of the proper motion distributions suggests that a sufficiently massive Segue 2 may be capable of creating a track / velocity misalignment near the impact, as has been seen in other streams due to the LMC \citep[e.g.,][]{erkal_modelling_2018, erkal_total_2019, koposov_piercing_2019, koposov_s5_2023, ji_kinematics_2021, shipp_measuring_2021, vasiliev_tango_2021}. 

The $5\times10^9\,M_\odot$ Hernquist and $10^9\,M_\odot$ Plummer models for Segue 2 do not affect the distributions as strongly, creating shoulders or double peaks instead of shifting the entire distribution. For example, both of these models create a shoulder near $\mu_{\phi_1} \sim 0.2$ in the Tracer Orbit synthetic streams. For the Fit Orbit synthetic streams (for which Segue 2's impact parameter is smaller), the $5\times10^9\,M_\odot$ Hernquist Segue 2 creates a slight shoulder around $\mu_{\phi_2} \sim -0.3$ and a second peak in $\mu_{\phi_1}$ near $\sim 0.2$ mas yr$^{-1}$, while the $10^9\,M_\odot$ Plummer Segue 2 does not. 

We also plot the proper motion distributions of the observed CPS (measured with respect to the \texttt{Cetus-Palca-T21} track) in gray. We stress that any comparison between the models and data in this region of the stream is subject to unknown uncertainties (see Section \ref{ssubsec:seg2_mass}), and note that there are only 35 observed stars in this region compared to $\approx$150 in the synthetic streams. However, as a proof of concept, the observed proper motion distributions have a shoulder in $\mu_{\phi_2}$ and a double-peak in $\mu_{\phi_1}$ reminiscent of those seen in the synthetic streams, suggesting that the velocity distribution of stars near the flyby location can distinguish between Segue 2 mass profiles. In particular, the proper motion distributions could help break the degeneracy in stream heating between a $10^9\,M_\odot$ Plummer and $5\times10^9\,M_\odot$ Hernquist profile. Breaking this degeneracy would provide insight on both the level of tidal stripping experienced by Segue 2 and whether DM cores can form in the absence of strong supernova feedback.

\section{Discussion} \label{sec:disc}

\subsection{Sensitivity of Our Results to Segue 2's Orbit}\label{subsec:dep_seg2_orbit}

As we have only considered Segue 2's fiducial orbit in Sections \ref{subsec:perturbation} \& \ref{subsec:models}, it is worth asking to what extent our results depend on Segue 2's orbit. In Figure \ref{fig:mockstreams}, we consider synthetic streams made with both CPS orbit models, which results in two different encounter distances, timings, velocities, and angles (see the bottom portion of Table \ref{fig:flyby_stats}). The heating in the CPS caused by Segue 2 is much more sensitive to changes in Segue 2's mass (compare different lines within the same panel) than changes in the details of the encounter (compare the same Segue 2 mass within a row). 

Additionally, as the CPS is a thick stream, individual stream stars in a given encounter will have a range of impact parameters and relative velocities to Segue 2 that is similar to the range of impact parameters and relative velocities allowed by the uncertainty in Segue 2's orbit. Therefore, we expect that the magnitude of the perturbation in the CPS primarily probes Segue 2's mass and is insensitive to Segue 2's orbit within its allowed parameter space. 

\subsection{Limitations of Our Models} \label{subsec:caveats}

In this section, we discuss the outstanding limitations inherent in our choice of modeling approach. Specifically, we address: 
the effect of the MW and LMC masses in Section \ref{ssubsec:MWLMC_mass},
our choice of rigid galaxy potentials in Section \ref{ssubsec:rigid_pots}, and the effect of dynamical friction in Section \ref{ssubsec:DF}.

\subsubsection{Milky Way and LMC Mass}\label{ssubsec:MWLMC_mass}

\begin{table}[]
    \caption{
    Model parameters for our alternative potentials.
    } \label{tab:alt_gal_models}
    \centering
    \begin{tabular}{c c c}
        \hline
        \hline
         Parameter & Value & Unit\\
         \hline
         \multicolumn{3}{c}{Heavy MW} \\
         \hline
         $M_\mathrm{vir}$ & $1.5\times10^{12}$ & $M_\odot$\\
         $R_\mathrm{vir}$ & 300.79 & kpc\\
         $c_\mathrm{vir}$ & 9.56  & -\\
         $M_\mathrm{disk}$ & $5.5 \times 10^{10}$  & $M_\odot$\\
         $R_\mathrm{disk}$ & 3.5 & kpc \\
         $z_\mathrm{disk}$ & 0.53 & kpc \\
         $M_\mathrm{H,bulge}$ & $1\times10^{10}$ & $M_\odot$ \\
         $a_\mathrm{bulge}$ & 0.7  & kpc\\
         \hline
         \multicolumn{3}{c}{Heavy LMC} \\
         \hline
         $M_\mathrm{H}$ & $2.5\times10^{11}$  & $M_\odot$\\
         $a$ & 25.2 & kpc \\
         \hline
         \hline
    \end{tabular}
    \tablenotetext{}{NOTE - Quantities listed are the same as in Table \ref{tab:gal_models}.}
\end{table}

\begin{figure*}
    \centering
    \includegraphics[width=\textwidth]{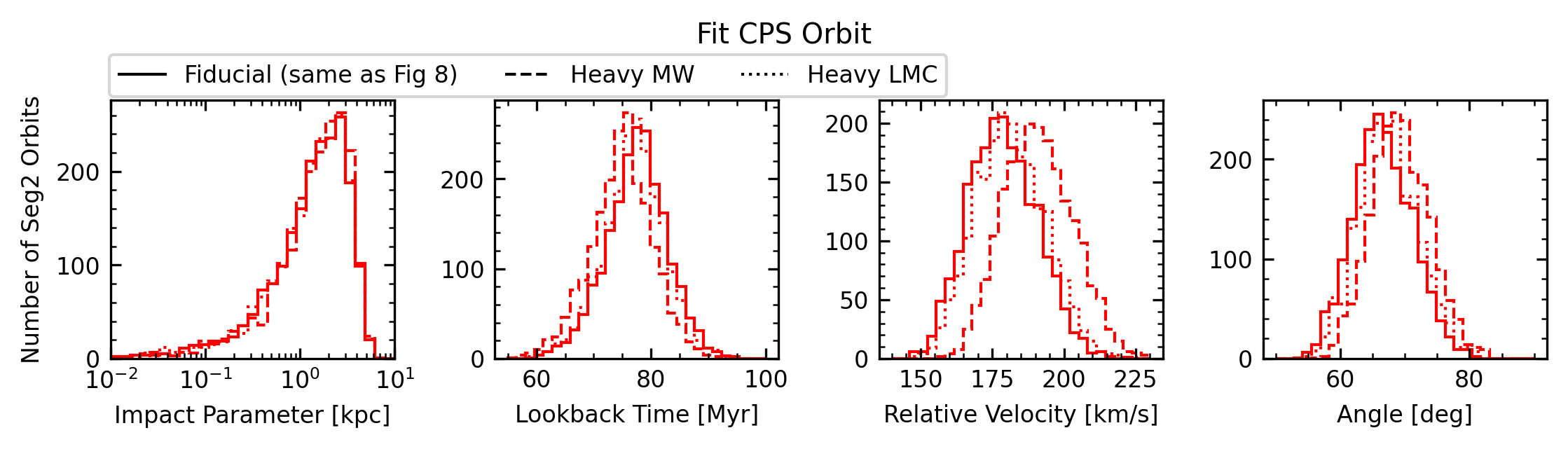}
    \includegraphics[width=\textwidth]{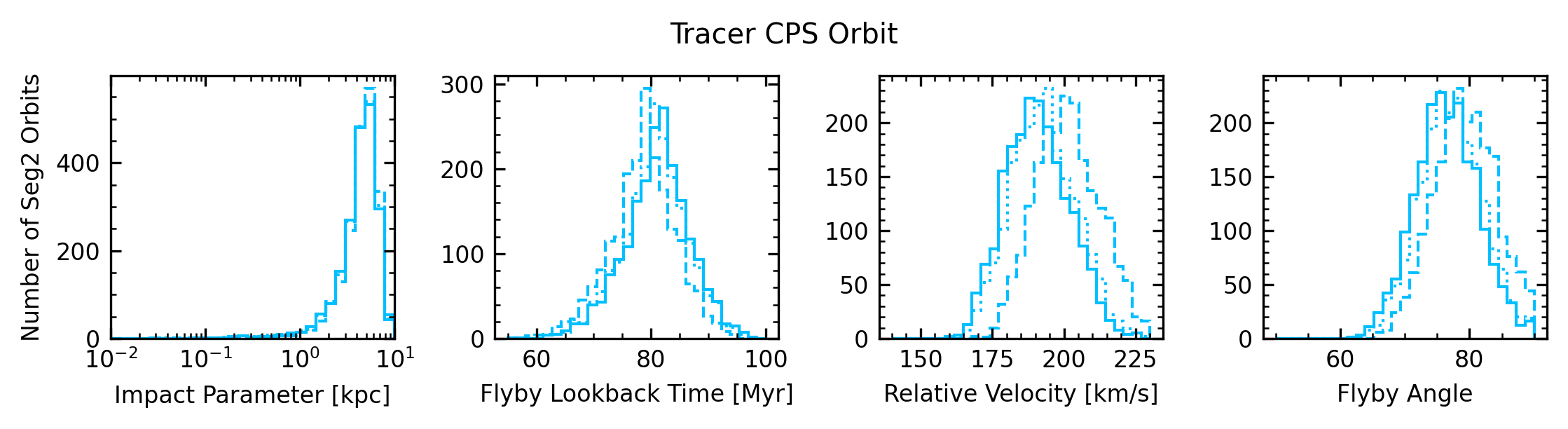}
    \caption{Comparison of the flyby parameter distributions in our Heavy MW (dashed lines) and Heavy LMC potentials (dotted lines) to those in Figure \ref{fig:flyby_stats} (solid lines). The probability of a close encounter, i.e. distribution of impact parameters, is unchanged by modifying the MW or LMC mass. The timing, relative velocity, and angle of the flyby are negligibly affected by changing the LMC mass. Changing the MW mass does not affect the shape of the distributions, but causes the encounter to happen at smaller lookback times, at a slightly higher relative velocity and angle. A higher MW mass causes the mean flyby timing, relative velocity, and angle to shift by $\approx$ 1$\sigma$ with respect to their values in the fiducial potential. 
    }
    \label{fig:masscomp}
\end{figure*}

\begin{table}[]
    \caption{
    Flyby parameters in our alternative potentials. 
    } \label{tab:masscomp_flybys}
    \centering
    \begin{tabular}{c c c c}
        \hline
        \hline
         Parameter & Measured w/  & Measured w/  & Unit \\
          & Fit Orbit & Tracer Orbit & \\
         \hline
         \multicolumn{4}{c}{Heavy MW} \\
         \hline
         Lookback Time & $74.9\pm5.4$ & $78.5\pm5.7$ & Myr\\
         Rel. Velocity & $191.0\pm12.1$ & $201.5\pm11.3$ & km s$^{-1}$\\
         Angle & $69.3\pm4.4$ & $79.2\pm4.8$ & deg \\
         \hline
         \multicolumn{4}{c}{Heavy LMC} \\
         \hline
         Lookback Time & $77.0\pm5.4$ & $80.6\pm5.7$ & Myr \\
         Rel. Velocity & $181.4\pm12.0$ & $192.9\pm11.2$ & km s$^{-1}$ \\
         Angle & $67.4\pm4.6$ & $77.1\pm4.8$ & deg \\
         \hline
         \hline
    \end{tabular}
    \tablenotetext{}{NOTE - The top section lists the mean flyby parameters for the Heavy MW, and the bottom section does the same for the Heavy LMC. Note that the impact parameter distributions are insensitive to the MW and LMC masses we test here (see Figure \ref{fig:masscomp}), so we omit the impact parameters for brevity.}
\end{table}

To determine the dependence of our results on the mass of the MW and LMC, we repeat the analyses of Sections \ref{sec:orbit} and \ref{subsec:flyby} with two alternative potentials, varying the mass of the Milky Way in one version and the mass of the LMC in the other. The model parameters for our ``Heavy MW'' and ``Heavy LMC'' are listed in Table \ref{tab:alt_gal_models}. The Heavy MW is the same as the MW2 model from \citet{patel_orbital_2020}, and the Heavy LMC is \citet{garavito-camargo_hunting_2019}'s LMC4.

The results of this experiment are shown in Figure \ref{fig:masscomp}, which shows the distributions of flyby parameters for the 2000 MC Segue 2 orbits similarly to Figure \ref{fig:flyby_stats}. The solid lines reproduce the distributions from Figure \ref{fig:flyby_stats} in our fiducial potential, while the dashed (dotted) lines show the distributions calculated with the Heavy MW (LMC). Looking at the left column of panels (impact parameter), it is clear that the probability of Segue 2's flyby with the CPS is unchanged by varying the mass of the MW or LMC. In other words, the CPS - Segue 2 interaction is recent enough that changes in the orbits of both objects in our alternative mass models are too small to prevent the encounter. A more massive LMC or MW does not affect the shape of the timing, relative velocity, or flyby angle distributions. Instead, a heavier LMC causes the flybys to occur slightly more recently, at a larger relative velocity and at larger angles, though these differences are negligible. A more massive MW causes the same effects, though the differences from the fiducial MW mass are more pronounced, at about the 1$\sigma$ level. Table \ref{tab:masscomp_flybys} contains a summary of the flyby timing and geometry in our alternative potentials. 

We note that these results imply that a \textit{less} massive MW would leave more time for gap formation in the CPS due to Segue 2's impact, which would relax the constraints on gap formation we derived in Section \ref{ssubsec:seg2_mass}.

\subsubsection{Deformation of the MW and LMC}\label{ssubsec:rigid_pots}

In this work, we have chosen to use rigid potentials for all the galaxies we consider. While we account for the bulk movement / reflex motion of the MW's center of mass in response to the LMC, our choice of rigid potentials fails to account for the shape and kinematic distortions to the halos of the MW and LMC owing to their interaction \citep{erkal_total_2019, erkal_equilibrium_2020, erkal_detection_2021, garavito-camargo_hunting_2019, garavito-camargo_quantifying_2021, cunningham_quantifying_2020, petersen_reflex_2020,  petersen_detection_2021, makarov_lmc_2023, vasiliev_dear_2024, chandra_all-sky_2024, sheng_uncovering_2024, yaaqib_radial_2024}. 

However, the CPS is currently far from the LMC, i.e., the CPS's apocenter is 38 (39) kpc for the Fit (Tracer) Orbit, compared to the LMC's present Galactocentric distance of 49 kpc \citep{pietrzynski_distance_2019}. In addition, while the CPS's orbit lies partially outside of 30 kpc (inside of which the MW halo remains relatively undisturbed by the LMC; \citealt{garavito-camargo_quantifying_2021}), the effects of MW halo distortion increase with Galactocentric radius \citep[e.g.,][]{laporte_response_2018, erkal_total_2019, erkal_equilibrium_2020, garavito-camargo_hunting_2019, garavito-camargo_quantifying_2021, vasiliev_effect_2023, sheng_uncovering_2024} and will remain small near the CPS apocenter. 
Moreover, we have demonstrated that our results are insensitive to changing the mass of the MW or LMC (see Section \ref{ssubsec:MWLMC_mass}), which will have a larger effect on the orbits of Segue 2 and the CPS than the detailed shape of the MW's halo. In short, we do not expect the shape of the MW's halo and/or presence of the LMC to change the probability that Segue 2 is impacting the CPS.

\subsubsection{Dynamical Friction}\label{ssubsec:DF}

In our simulations, we have neglected the effect of dynamical friction, which in general causes the orbits of objects within a host galaxy's DM halo to decay over time \citep{chandrasekhar_dynamical_1943}. To confirm that dynamical friction does not cause a significant change in the orbits of MW satellites over the timescales we consider in this work (the past 500 Myr), we compare the LMC's orbit over this time in our simulations to an orbit with dynamical friction from \citet{patel_orbital_2020}.\footnote{Note that while \citet{patel_orbital_2020} report the LMC's orbit in the presence of the SMC, we compare our simulations to an orbit that does not include the SMC for consistency.} Their orbit is computed in their MW1 potential (which we also use in this work; see Section \ref{subsec:gals}). By 500 Myr, the LMC's position in our simulation differs from that of \citet{patel_orbital_2020} by just $\approx 2$ kpc. Over the timescales in which Segue 2 interacts with the CPS (< 100 Myr ago), the LMC's position in the two simulations is effectively identical. The strength of the dynamical friction drag force scales as $M_{\rm{sat}}/v_{\rm{sat}}^2$ \citep{chandrasekhar_dynamical_1943}, so the drag force on our most massive Segue 2 model will be $\approx 3.6$ times weaker than the drag on the LMC at present. Therefore, neglecting dynamical friction in our simulations has no bearing on our results. 

\subsection{Limitations of Our Data}\label{subsec:data_lims}

While our CPS dataset is sufficient for fitting the stream's orbit, it has several limitations that prevent careful comparisons to our models of the CPS - Segue 2 interaction. Specifically, there are two major shortcomings discussed in Section \ref{sec:segue2}, which we summarize below before outlining the additional data that will be needed to perform accurate comparisons with our models:

\textit{Too few radial velocity measurements ahead of Segue 2's impact point:} Our 6-D dataset contains only one star ahead of Segue 2's impact point (bottom panel of Figure \ref{fig:star_perturb}) where the stream is expected to be most strongly affected by Segue 2 (see Figures \ref{fig:model_perturb}). Without 6-D data in the Impact Region of the stream, we cannot search for Segue 2's effect on the CPS by integrating the orbits of stars in the affected portion of the stream.

\textit{Reliance on velocity clustering for sample selection:} During the selection process for our CPS catalog (see Section \ref{sec:obs}), we employed a series of proper motion cuts to identify stars that have on-sky motions consistent with CPS membership. These proper motion cuts are independent of $\phi_1$, implying that any stars which were scattered more than 0.4 mas yr$^{-1}$ away from the CPS's proper motion track would be excluded from our observational dataset. This fundamentally limits comparisons of the CPS's velocity distribution in the vicinity of the flyby to our models, especially as some Segue 2 models we consider are capable of scattering CPS stars outside of our proper motion cuts (see Figure \ref{fig:vDists}).

Moving forward, both shortcomings can be overcome by spectroscopically observing as many stars in an area surrounding the flyby (roughly $\phi_1 > 0^\circ$, $|\phi_2| < 15^\circ$) as possible, providing both radial velocity measurements and chemical abundances for these stars. Greater radial velocity coverage of the CPS will increase the completeness of our 6D sample and improve the accuracy of the measured CPS velocity dispersion as a function of $\phi_1$, as spectroscopic radial velocities typically have smaller errors than proper motions at $\sim$35 kpc.\footnote{The mean radial velocity error in our CPS catalog is 2.0 km s$^{-1}$ compared to a mean proper motion error of 0.14 mas yr$^{-1}$, which is 23.3 km s$^{-1}$ converted to a linear velocity assuming a distance of 35 kpc.} 

Chemical abundances will provide an additional criterion for determining if a star belongs to the CPS, increasing the purity of the CPS member catalog and improving knowledge of the selection function. These data will also enable a search for stars that lie farther from the CPS's velocity track (i.e. outside of the proper motion cuts used in this work) which share the stream's chemical composition. In short, stellar chemistry would increase both the completeness and purity of our CPS catalog, enabling accurate measurements of the CPS' velocity distribution and comparisons to our models. 

Fortunately, there is a growing effort to conduct industrial-scale spectroscopic follow-up of stellar streams \citep[e.g.,][]{bonaca_stellar_2025} using facilities such as  DESI \citep{cooper_overview_2023}, WEAVE \citep{jin_wide-field_2024}, 4MOST \citep{de_jong_4most_2019}, PFS \citep{takada_extragalactic_2014}, and Via\footnote{https://via-project.org/}. The CPS - Segue 2 interaction presents an intriguing target for these surveys, and we look forward to using large-scale spectroscopic observations of the CPS for future work.

\subsection{Triangulum / Pisces Stream Candidates} \label{subsec:tripsc}

\begin{figure*}
    \centering
    \includegraphics[width=\textwidth]{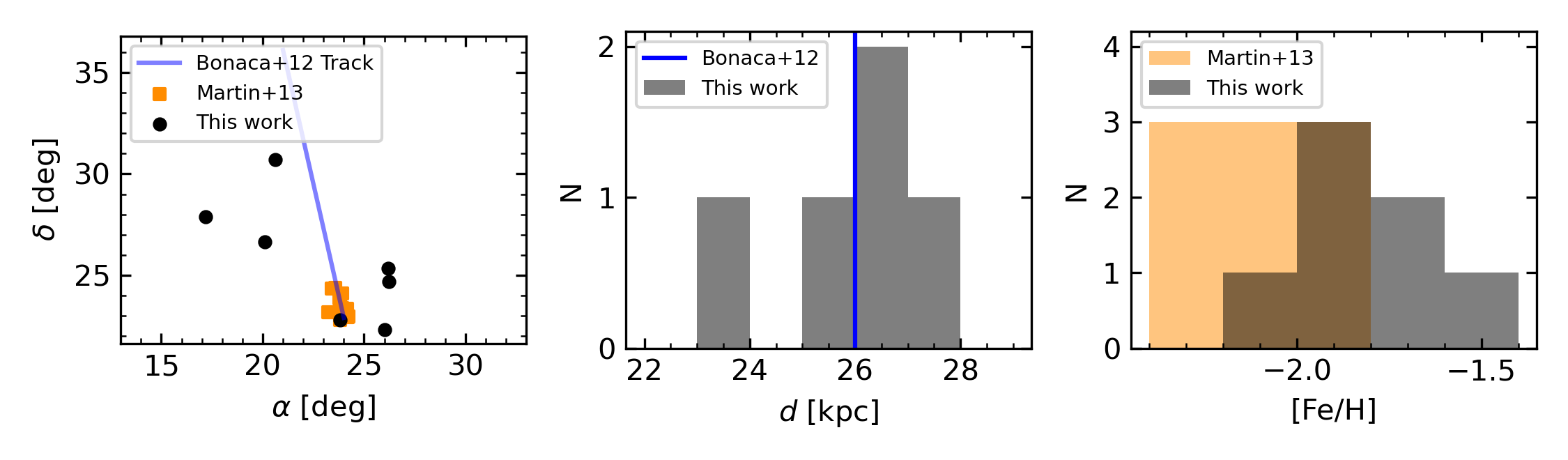}
    \caption{Comparison of our Tri/Psc candidates to the \texttt{Tri-Pis-B12} track and the \citet{martin_kinematic_2013} Tri/Psc catalog. The left panel shows the sky positions of our stars (black points), the \citet{martin_kinematic_2013} stars (orange squares), and the \texttt{Tri-Pis-B12} track (blue line). One of our stars is included in the \citet{martin_kinematic_2013} sample. The remaining stars cluster around the track with a large spread. In the center panel, we show the distance distribution of our candidates. \citet{bonaca_cold_2012} measured the distance of Tri/Psc to be 26 kpc (indicated by the blue line). The right panel compares the metallicity distribution of our candidates (black) to that of the \citet{martin_kinematic_2013} stars (orange). While our candidates are spread out with respect to the stream track, they are consistent with Tri/Psc in spatial location, metallicity, and radial velocity (see Figure \ref{fig:rv}), suggesting they may correspond to an envelope or cocoon structure.}
    \label{fig:tripsc}
\end{figure*}

Here, we conduct a further investigation into the group of seven stars identified in the purple box in Figure \ref{fig:rv} as having radial velocities and longitudes consistent with the Tri/Psc stream. Figure \ref{fig:tripsc} compares the on-sky positions of these stars to the \texttt{Tri-Pis-B12} track from \citet{bonaca_cold_2012} published in \texttt{galstreams} and to a catalog of spectroscopic Tri/Psc members published by \citet{martin_kinematic_2013}. One of the \citet{martin_kinematic_2013} stars is present in our candidate list. While the remaining six candidates are $\gtrsim2^\circ$ from the stream track, their distances and metallicities are roughly consistent with constraints on Tri/Psc from \citet{bonaca_cold_2012} and \citet{martin_kinematic_2013} respectively. 

Therefore, we tentatively propose that the six candidates not previously identified as Tri/Psc members may be part of a wider ``cocoon'' component of the stream, similar to those found around GD-1 \citep{malhan_butterfly_2019, valluri_gd-1_2024} and Jhelum \citep{bonaca_multiple_2019, awad_swarming_2024}. Such a hot envelope around Tri/Psc would be consistent with the picture that its globular cluster progenitor once belonged to the dwarf galaxy progenitor of the CPS as proposed by \citet{bonaca_orbital_2021} (see also \citetalias{thomas_cetus-palca_2022}; \citealt{yuan_complexity_2022}). In this scenario, the hot envelope could be created via preprocessing of the globular cluster inside the Cetus-Palca dwarf \citep[e.g.,][]{carlberg_globular_2018} and could be used to constrain the density profile of the Cetus-Palca dwarf's DM halo \citep[e.g.,][]{malhan_probing_2021}. Additional study will be needed to confirm the association between these stars and the cold component of the stream. 

\section{Conclusions} \label{sec:con}

In this paper, we have used the observed kinematics of the CPS and Segue 2 to discover and study the interaction between these two members of the MW halo. Our main findings are summarized below. 

\begin{itemize}
    \item \textbf{Beginning with the catalog of \citetalias{thomas_cetus-palca_2022}, we identify a relatively pure sample of 403 RGB and BHB stars belonging to the CPS. 39 of these stars have spectroscopic radial velocities measured with SEGUE or H3 (Figures \ref{fig:sel}, \ref{fig:track_sel}, \& \ref{fig:rv}).} Our sample is consistent with other detections of the main CPS wrap \citep{yuan_revealing_2019, yuan_complexity_2022}. However, we refrain from searching for other components of the CPS identified by these authors. 
    
    \item \textbf{We employ two methods for inferring the orbit of the CPS, fitting the orbit assuming that it is roughly described by the stream track (``Fit Orbit''), and by finding the orbit of an exemplary member star (``Tracer Orbit''; Figure \ref{fig:repstar}).} The resulting orbits (Figures \ref{fig:orbits} \& \ref{fig:sky_orbit}) trace the stream track well, and are in broad agreement with the findings of \citet{chang_is_2020} and \citetalias{thomas_cetus-palca_2022}. 

    \item \textbf{By considering 2000 possible orbits of Segue 2 based on uncertainties in its distance and velocity vector, we find that Segue 2 directly impacted (passed within the stream's $2\sigma$ width) the CPS 77$\pm$5 Myr ago (Figures \ref{fig:mc_orbits} \& \ref{fig:flyby_stats}).} When using our Fit CPS Orbit as a proxy for the stream location, all but two of the possible Segue 2 orbits pass within the CPS's 2$\sigma$ width (5.84 kpc), and the impact occurs 77$\pm$5 Myr ago at a relative velocity of $179\pm12$ km and an angle of $67\pm5$ degrees. With our Tracer CPS Orbit, 77.16\% of the Segue 2 orbits pass within the same distance to the stream, and the impact occurs 81$\pm$6 Myr ago at a relative velocity of $191\pm11$ km s$^{-1}$ and an angle of $76\pm5$ degrees. These results are insensitive to the LMC mass assumed, but raising the MW mass by 50\% makes the impact occur $\sim$ 4 Myr earlier (Figure \ref{fig:masscomp}). This is the first known interaction between a stream and a UFD. Crucially, Segue 2's velocity and position are known \textit{a priori}, so this interaction can be used to test our theoretical understanding of stream perturbations. 

    \item \textbf{By integrating the orbits of our 6-D sample of stars, we find that Segue 2 is expected to perturb the velocities of stream stars by up to $\sim 20$ km s$^{-1}$ assuming Segue 2 has a total mass of $\mathbf{10^9}$ $\mathbf{M_\odot}$ and is described by a Plummer profile with a scale radius of 0.865 kpc (Figure \ref{fig:star_perturb}).} The magnitude of this perturbation is similar to the intrinsic velocity dispersion of the stream. The stars with the smallest impact parameters / largest perturbations from Segue 2 are expected to be in $20^\circ < \phi_1 < 40^\circ$ at the present day (Figure \ref{fig:model_perturb}). Unfortunately, our 6-D observational data only covers one side of the flyby well, as Segue 2 passes barely through the leading edge of our 6-D stars. Additional radial velocity observations of stars at $\phi_1 > 0^\circ$ and a selection strategy that takes into account potential heating from Segue 2 are needed to confirm the presence or absence of a perturbation in the CPS from Segue 2. 

    \item \textbf{Using synthetic stream models, we show that the primary signature of Segue 2's flyby is an increase in all velocity dispersion components of the CPS at $\mathbf{\phi_1 \gtrsim 10^\circ}$, peaking near the Impact Region ($20^\circ < \phi_1 < 40^\circ$; Figure \ref{fig:mockstreams}). If such a perturbation is detected, it would place a lower limit on Segue 2's mass within $\sim6$ kpc of $\mathbf{> 10^9}$ $\mathbf{M_\odot}$.} Below this mass, Segue 2 is not expected to produce detectable heating, i.e. velocity kicks imparted to stream stars by Segue 2 are smaller than the stream's width in velocity space. We expect this result is insensitive to the uncertainty in Segue 2's orbit. As such, a conclusive detection or non-detection of heating in the CPS from Segue 2 would provide a measurement or upper limit to Segue 2's halo mass. 

    \item \textbf{The density and proper motion dispersions of our synthetic stream models have high-level inconsistencies with the observed CPS if Segue 2's mass is $\mathbf{10^{10}\,M_\odot}$ within $\sim6$ kpc (Figure \ref{fig:mockstreams}).} Therefore, we tentatively place an upper limit on Segue 2's mass within 6 kpc of $10^{10}\,M_\odot$ assuming Segue 2 is well described by a Plummer sphere. A more detailed future modeling effort coupled with a more careful search for CPS stars outside of the established velocity track will be required to confirm or refute this result. 

    \item \textbf{We should not expect to observe a gap in the CPS from Segue 2's impact at present.} The earliest encounter allowed by our search of the error space of Segue 2's orbit occurs less than 100 Myr ago, which leaves insufficient time for a gap to form, even if Segue 2's mass is $10^{10}\,M_\odot$.

    \item \textbf{
    Given that Segue 2's mass within its half-light radius is constrained via the internal dynamics of its stars, the slope of Segue 2's density profile can be constrained by a measurement of its mass at $\sim 6$ kpc from the kinematics of CPS stars near the impact (Figure \ref{fig:m_profs}).} In particular, if a perturbation to the CPS consistent with $M_{\mathrm{Seg2}}(<6 \,\rm{kpc}) = 5\times10^9\,M_\odot$ is found, Segue 2's inner density profile must be more shallow than an NFW or Hernquist profile. 
    
    \item \textbf{The proper motion distribution of stars in the Impact Region ($20^\circ < \phi_1 < 40^\circ$) is mildly sensitive to Segue 2's extended mass distribution (Figure \ref{fig:vDists}).} This suggests the shape of the velocity distribution may also be useful for constraining Segue 2's density profile and degree of tidal truncation as future observations increase the spectroscopic coverage of this region of the stream. 

    \item \textbf{We report the serendipitous discovery of a potential envelope or cocoon around the Tri/Psc stream (Figure \ref{fig:tripsc}).} In the population of \citetalias{thomas_cetus-palca_2022} stars that we excluded from the CPS, we found seven stars with radial velocities and metallicities consistent with the Tri/Psc stream. These stars cluster around the 3-D position of Tri/Psc with a spread of several degrees, much larger than the width of the cold stream ($\approx 0.2^\circ$; \citealt{bonaca_cold_2012}). If these stars are confirmed as a hot component of Tri/Psc, this finding would lend evidence to the scenario proposed by \citet{bonaca_orbital_2021} in which Tri/Psc formed from a globular cluster belonging to the CPS's progenitor.
    
\end{itemize}

Moving forward, 
increasingly deep spectroscopic surveys (e.g.,, DESI \citep{cooper_overview_2023}, WEAVE \citep{jin_wide-field_2024}, 4MOST \citep{de_jong_4most_2019}, PFS \citep{takada_extragalactic_2014}, and Via\footnote{https://via-project.org/}) will provide a more complete census and greater radial velocity coverage of the CPS. Together with future modeling efforts based on these datasets, these observations will serve as an excellent proving ground for the basic predictions made here. Ideally, observations should aim to measure the velocity dispersion of the CPS as a function of longitude to the $\approx$1 km s$^{-1}$ level with a well-understood selection function. Our models predict that the dispersion of the CPS at $\phi_1\gtrsim10^\circ$ will be $\sim 60$ km s$^{-1}$ compared to $\sim 20$ km s$^{-1}$ at $\phi_1\lesssim0^\circ$ if Segue 2's mass within $\sim6$ kpc is $5\times10^9\,M_\odot$. Conversely, if the CPS's dispersion is constant with longitude, an upper limit of $10^9\,M_\odot$ can be placed on Segue 2's mass enclosed within $\sim6$ kpc.

The CPS - Segue 2 interaction is a rare example of a close encounter between a stellar stream and a known MW satellite galaxy. This scenario is an ideal test bed for models designed to infer the properties of an unknown dark perturber from a stream perturbation, such as those developed by  \citet{erkal_properties_2015}, \citet{bonaca_spur_2019}, and \citet{hilmi_inferring_2024}. While a thick dwarf galaxy stream like the CPS is inherently less sensitive to Segue 2 than a thin globular cluster stream, verifying that these models can recover Segue 2's position, velocity, mass, and size from CPS perturbations will confirm their ability to discover the low-mass dark halos predicted by $\Lambda$CDM from perturbations in colder streams. 

Additionally, this interaction offers a unique probe of the DM halo of one of the most DM-dominated galaxies known. 
A constraint on Segue 2's halo mass within $\sim 6$ kpc from its influence on the CPS will provide a new mass measurement for a UFD halo at a larger radius than is possible with its internal dynamics, placing a rare data point on the faint end of the stellar mass - halo mass relation. Specifically, idealized simulations \citep{bland-hawthorn_ultrafaint_2015} and studies of the MW satellite population \citep{jethwa_upper_2018, kim_missing_2018, nadler_milky_2020} suggest UFDs reside in halos of masses $< 10^9\,M_\odot$, while cosmological simulations \citep{shen_baryon_2014, wheeler_sweating_2015, sawala_chosen_2016, jeon_connecting_2017, jeon_highly_2021, jeon_role_2021, munshi_quantifying_2021} predict UFD halo masses of $\approx 10^9\,M_\odot$ with roughly half a dex of scatter. Therefore, placing an upper limit on Segue 2's mass of $10^9\,M_\odot$ would suggest that these different methods of predicting the DM masses of the faintest galaxies are consistent. Conversely, if Segue 2's mass is measured to be $> 10^9\,M_\odot$ via a confirmed perturbation in the CPS, this would suggest cosmological simulations provide a more accurate picture of UFD formation. Constraining Segue 2's mass will provide clarity on the many theoretical predictions of UFD halo masses and inform the design of future studies on this topic.

In addition to its mass, Segue 2's DM density profile can be constrained by: 1) determining its slope from  existing estimates of Segue 2's mass within its half-light radius \citep{belokurov_discovery_2009, kirby_segue_2013} combined with a second data point at $\sim6$ kpc from Segue 2's influence on the CPS; and 2) examining the velocity distribution of CPS stars near the impact site. Measuring Segue 2's DM density profile would: 1) clarify its degree of tidal truncation, solving the longstanding debate on whether it is strongly affected by the MW's tides \citep{belokurov_discovery_2009, kirby_segue_2013, simon_faintest_2019, pace_proper_2022}; and 2) provide a critical test of the core-cusp problem by establishing whether a low-mass DM halo that should be minimally affected by baryonic physics can form a core, without relying solely on uncertain assumptions inherent in modeling Segue 2's internal dynamics. If Segue 2 is found to be cored, this would therefore suggest some level of self-interactions between its DM particles \citep[e.g.,][]{vogelsberger_dwarf_2014}.

In short, further study of the CPS - Segue 2 interaction will shed light on the formation process of UFDs and the path forward for constraining the nature of DM at the smallest galactic scales. 

\section{Acknowledgments}

We thank the anonymous referee for a thorough and insightful report, which improved the presentation of our results and the clarity of the manuscript. HRF would like to thank Peter Behroozi, Emily Cunningham, Kathryne Daniel, Arjun Dey, Kathryn Johnston, and Monica Valluri for helpful discussions. 

We respectfully acknowledge the University of Arizona is on the land and territories of Indigenous peoples. Today, Arizona is home to 22 federally recognized tribes, with Tucson being home to the O’odham and the Yaqui. Committed to diversity and inclusion, the University strives to build sustainable relationships with sovereign Native Nations and Indigenous communities through education offerings, partnerships, and community service.

HRF and GB are supported by NSF CAREER AST-1941096. EP is supported by NASA through Hubble Fellowship grant \# HST-HF2-51540.001-A, awarded by the Space Telescope Science Institute (STScI). STScI is operated by the Association of Universities for Research in Astronomy, Incorporated, under NASA contract NAS5-26555. 

The H3 Survey is funded in part by NSF award AST-2107253. This work has made use of data from the European Space Agency (ESA) mission
{\it Gaia} (\url{https://www.cosmos.esa.int/gaia}), processed by the {\it Gaia}
Data Processing and Analysis Consortium (DPAC,
\url{https://www.cosmos.esa.int/web/gaia/dpac/consortium}). Funding for the DPAC
has been provided by national institutions, in particular the institutions
participating in the {\it Gaia} Multilateral Agreement. We have made extensive use of NASA's Astrophysics Data System and the arXiv pre-print service in the preparation of this paper. 

%

\vspace{5mm}
\facilities{\textit{Gaia}, MMT/Hectochelle, Sloan}


\software{
\texttt{Astropy} \citep{astropy:2013, astropy:2018, astropy:2022};
\texttt{Gala} \citep{price-whelan_gala_2017, price-whelan_adrngala_2023};
\texttt{galstreams} \citep{mateu_galstreams_2023};
\texttt{Matplotlib} \citep{hunter_matplotlib_2007};
\texttt{NumPy} \citep{harris_array_2020};
\texttt{pandas} \citep{mckinney_pandas_2010, reback_pandas_2020}; 
\texttt{SciPy} \citep{virtanen_scipy_2020};
\texttt{TOPCAT} \citep{taylor_topcat_2005};
          }

\bibliography{references}{}
\bibliographystyle{aasjournal}



\end{document}